# Programming Metamorphic Algorithms

## An Experiment in Type-Driven Algorithm Design

Hsiang-Shang Ko[a]

a    Institute of Information Science, Academia Sinica, Taiwan

**Abstract**    In *dependently typed programming*, proofs of basic, structural properties can be embedded implicitly into programs and do not need to be written explicitly. Besides saving the effort of writing separate proofs, a most distinguishing and fascinating aspect of dependently typed programming is that it makes the idea of *interactive type-driven development* much more powerful, where expressive type information becomes useful hints that help the programmer to complete a program. There have not been many attempts at exploiting the full potential of the idea, though. As a departure from the usual properties dealt with in dependently typed programming, and as a demonstration that the idea of interactive type-driven development has more potential to be discovered, we conduct an experiment in 'type-driven algorithm design': we develop algorithms from their specifications encoded in sophisticated types, to see how useful the hints provided by a type-aware interactive development environment can be. The algorithmic problem we choose is *metamorphisms*, whose definitional behaviour is consuming a data structure to compute an intermediate value and then producing a codata structure from that value, but there are other ways to compute metamorphisms. We develop Gibbons's streaming algorithm and Nakano's jigsaw model in the interactive development environment provided by the dependently typed language Agda, turning intuitive ideas about these algorithms into formal conditions and programs that are correct by construction.

# The Art, Science, and Engineering of Programming

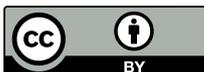

## 1 Interactive Type-Driven Development

With the use of indexed (co-)datatypes [8, 29], *dependently typed programming* [18, 22] has been successful in offering 'intrinsic' guarantees about basic properties whose proofs closely follow program structures and can be implicitly embedded into programs. Typical examples include keeping indices within bounds [11], representing well-scoped and well-typed syntax trees of embedded languages [2], and tracking the available effects/resources used in programs [6]. All these examples involve some form of constraints: bounds, typing, or available resources. In dependently typed programming, these constraints are embedded into the respective datatypes (of indices, syntax trees, and program operations) so that the type system can help the programmer to enforce the constraints without writing additional proofs. But this kind of guarantee is also achievable 'extrinsically' with proof assistants, where proofs are written explicitly and separately from programs — we know how to prove theorems saying that indices are always within bounds or that a program transformation always produces well-typed programs, and it is not hard to prove these theorems formally, especially with proof assistants that have decent automation [3, 24]. What distinguishes dependently typed programming from other methodologies is that it makes the idea of *interactive type-driven development* much more powerful: the more properties we encode intrinsically into types, the more hints the type-checker can offer the programmer during an interactive development process, so that the more heavyweight typing becomes an aid rather than a burden [14, 19, 20].

A natural follow-up question is: how well does interactive type-driven development scale? For instance, if we encode a complete specification of an algorithmic problem in a type, will the hints provided by the type-checker be useful enough to help us with the formal construction of an algorithm (about which we may already have an intuitive idea but have not worked out the formal detail)? Thankfully, we now have dependently typed languages like Agda [16, 25, 28] and Idris [5, 7] whose interactive development environment is mature enough for doing some experiments — that is, trying to program some algorithms interactively with the type-checker. In this paper we will use Agda 2.6.1 with Standard Library version 1.3. The code in this paper is available [13].

The ideal reader of this paper is someone who has experience in Agda programming (for example, who has worked through Part 1 of *Programming Language Foundations in Agda* [16]) and wishes to see some algorithmic examples. On the other hand, for a reader with some basic knowledge about functional programming and dependent types (e.g., dependent function and pair types and indexed datatypes), the paper will provide necessary explanations about Agda-specific details so that the reader can at least get a sense of what it feels like to program interactively in Agda.

To make our experiment meaningful and more likely to succeed, we had better choose an algorithmic problem that is not too trivial and also not too complicated. One such problem that comes to mind is metamorphisms.

## 2  Metamorphisms

A *metamorphism* [10] consumes a data structure to compute an intermediate value and then produces a new data structure using the intermediate value as the seed. Following Gibbons [10], we will focus on *list metamorphisms*, i.e., metamorphisms consuming and producing lists. Two of the examples of list metamorphisms given by Gibbons are

- *Base conversion for fractions*: A list of digits representing a fractional number in a particular base (e.g., $0.625_{10}$) can be converted to another list of digits in a different base (e.g., $0.101_2$). The conversion is a metamorphism because we can consume an input list of digits to compute the value represented by the list, and then produce an output list of digits representing the same value. Note that even when the input list is finite, the output list may have to be infinite — for example, $0.1_3 = 0.333\ldots_{10}$.
- *Heapsort*: An input list of numbers is consumed by pushing every element into a min-heap (priority queue), from which we then pop all the elements, from the smallest to the largest, producing a sorted output list.

We will use these two examples to provide some intuition about metamorphisms, but the examples will not be our focus. Instead, we will mainly be working on a number of generic algorithms that compute metamorphisms [10, 23]. Knowing the existence of these algorithms, our experiment will have a clearer direction to follow and be more likely to succeed. However, rather than merely implementing them in Agda, we should try to reinvent as much of the algorithms as possible, so that we can feel what it is like to get from some algorithmic intuition and a specification to a formal algorithm with the help of Agda, and perhaps gain a different and/or better understanding of these algorithms.

Let us first say more precisely what metamorphisms are. Formally, a metamorphism is a *fold* followed by an *unfold*, the former consuming a finite data structure and the latter producing a potentially infinite codata structure.

**Lists for consumption**   For list metamorphisms, the inputs to be consumed are the standard finite lists:[1]

**data** List $(A : \mathrm{Set}) : \mathrm{Set}$ **where**
   [ ]   : List $A$
   $\_::\_ : A \to \mathrm{List}\,A \to \mathrm{List}\,A$

The *foldr* operator subsumes the elements (of type $A$) of a list into a state (of type $S$) using a 'right algebra' $(\lhd) : A \to S \to S$ and an initial/empty state $e : S$:[2]

---

[1] In Agda, a name with underscores like $\_::\_$ can be used as an operator, and the underscores indicate where the arguments go. However, some binary infix operators like $\_\lhd\_$ are written in Haskell syntax like $(\lhd)$ (which is not syntactically correct in Agda). All subsequent footnotes will be about Agda-specific details, and can be safely skipped by Agda programmers.

[2] In the type of a function, arguments wrapped in curly brackets are implicit, and can be left out (if they are inferable by Agda) when applying the function.

$$foldr : \{A\ S : \mathsf{Set}\} \rightarrow (A \rightarrow S \rightarrow S) \rightarrow S \rightarrow \mathsf{List}\ A \rightarrow S$$
$$foldr\ (\lhd)\ e\ [\ ] \qquad = e$$
$$foldr\ (\lhd)\ e\ (a :: as) = a \lhd foldr\ f\ e\ as$$

With *foldr*, a list is consumed from the right. Dually, the *foldl* operator consumes a list from the left using a 'left algebra' $(\rhd) : S \rightarrow A \rightarrow S$:

$$foldl : \{A\ S : \mathsf{Set}\} \rightarrow (S \rightarrow A \rightarrow S) \rightarrow S \rightarrow \mathsf{List}\ A \rightarrow S$$
$$foldl\ (\rhd)\ e\ [\ ] \qquad = e$$
$$foldl\ (\rhd)\ e\ (a :: as) = foldl\ (\rhd)\ (e \rhd a)\ as$$

A list metamorphism can use either *foldr* or *foldl* in its consuming part, and we will see both kinds in the paper. We will refer to a list metamorphism using *foldr* as a 'right metamorphism', and one using *foldl* as a 'left metamorphism'.

**Colists for production**   For the producing part of list metamorphisms, where we need to produce potentially infinite lists, in a total language like Agda we can no longer use List, whose elements are necessarily finite; instead, we should switch to a *codatatype* of *colists*, which are potentially infinite. Dual to a datatype, which is defined by all the possible ways to *construct* its elements, a codatatype is defined by all the possible ways to *deconstruct* or *observe* its elements. For a colist, only one observation is possible: exposing the colist's outermost structure, which is either empty or a pair of a head element and a tail colist. In Agda this can be expressed as a coinductive record type CoList with only one field, which is a sum structure that is either empty or non-empty; for cosmetic reasons, this sum structure is defined (mutually recursively with CoList) as a datatype CoListF:[3]

```
mutual
  record CoList (B : Set) : Set where
    coinductive
    field
      decon : CoListF B
  data CoListF (B : Set) : Set where
    [ ]   : CoListF B
    _::_ : B → CoList B → CoListF B
  open CoList
```

Note that decon denotes a function of type $\{B : \mathsf{Set}\} \rightarrow \mathsf{CoList}\ B \rightarrow \mathsf{CoListF}\ B$, and plays the role of the deconstructor of CoList. Now we can define the standard *unfoldr*

---

Agda provides some ways to shorten type signatures. For one, named arguments of the same type can be declared together: here, for example, '$\{A\ S : \mathsf{Set}\} \rightarrow$' is an abbreviation of '$\{A : \mathsf{Set}\} \rightarrow \{S : \mathsf{Set}\} \rightarrow$'. Also, in the latter type fragment, the arrow between typed arguments can be skipped, resulting in another valid form '$\{A : \mathsf{Set}\}\ \{S : \mathsf{Set}\} \rightarrow$'.

[3] Agda supports constructor overloading, so we are allowed to reuse [ ] and _::_ as the names of the constructors of CoListF.

operator, which uses a coalgebra $g : S \to \mathsf{Maybe}\,(B \times S)$ to unfold a colist from a given state:[4]

$unfoldr : \{B\,S : \mathsf{Set}\} \to (S \to \mathsf{Maybe}\,(B \times S)) \to S \to \mathsf{CoList}\,B$
$\mathsf{decon}\,(unfoldr\,g\,s)\;\textbf{with}\,g\,s$
$\mathsf{decon}\,(unfoldr\,g\,s)\;|\;\mathsf{nothing}\quad = [\,]$
$\mathsf{decon}\,(unfoldr\,g\,s)\;|\;\mathsf{just}\,(b,\,s') = b :: unfoldr\,g\,s'$

The operator is defined with copattern matching [1]: To define $unfoldr\,g\,s$, which is a colist, we need to specify what will be produced if we deconstruct it, that is, the result of $\mathsf{decon}\,(unfoldr\,g\,s)$. This result depends on whether $g$ can produce anything from $s$, so, using the **with** construct, we introduce $g\,s$ as an additional 'argument', on which we can then perform a pattern match. If $g\,s$ is nothing, then the resulting colist will be empty — that is, $\mathsf{decon}\,(unfoldr\,g\,s)$ will compute to $[\,]$; otherwise, $g\,s$ is just $(b,\,s')$ for some $b$ and $s'$, and the resulting colist will have $b$ as its head and $unfoldr\,g\,s'$ as its tail — that is, $\mathsf{decon}\,(unfoldr\,g\,s)$ will compute to $b :: unfoldr\,g\,s'$.

To be more concrete, let us describe our two examples — base conversion for fractions and heapsort — explicitly as metamorphisms. We look at heapsort first as it is simpler.

**Heapsort**  Let Heap be an abstract type of min-heaps on some totally ordered set Val, equipped with three operations

$empty\quad : \mathsf{Heap}$
$push\quad : \mathsf{Val} \to \mathsf{Heap} \to \mathsf{Heap}$
$popMin : \mathsf{Heap} \to \mathsf{Maybe}\,(\mathsf{Val} \times \mathsf{Heap})$

where $empty$ is the empty heap, $push$ adds an element into a heap, and $popMin$ returns a minimum element and the rest of the input heap if and only if the input heap is non-empty. Then heapsort can be directly described as a right metamorphism:

$unfoldr\,popMin \circ foldr\,push\,empty$

(In this case the length of the output colist is always finite. This fact is not captured by the type, which, however, should cover other cases where the length of the output colist can be infinite, such as base conversion for fractions.)

**Base conversion for fractions**  Suppose that the input and output bases are $b_i : \mathbb{N}$ and $b_o : \mathbb{N}$ — in $0.625_{10} = 0.101_2$, for example, $b_i = 10$ and $b_o = 2$. We represent fractions as (co)lists of digits (of type $\mathbb{N}$) starting from the most significant digit — for example, $0.625$ is represented as $6 :: 2 :: 5 :: [\,]$. Gibbons [10, Section 4.2] gives a more complete story, where base conversion for fractions is first described as a right metamorphism with simple states (consisting of only an accumulator), and then transformed to a left metamorphism with more complex states. To make the story short here, we go directly for a left metamorphism

---

[4] The datatype $\mathsf{Maybe}\,X$ has two constructors, $\mathsf{nothing} : \mathsf{Maybe}\,X$ and $\mathsf{just} : X \to \mathsf{Maybe}\,X$.

$unfoldr\ g_C \circ foldl\ (\triangleright_C)\ e_C$

where the state type is $S_C = \mathbb{Q} \times \mathbb{Q} \times \mathbb{Q}$, which are triples of rationals of the form $(v,\ w_i,\ w_o)$ where $v$ is an accumulator, $w_i$ the weight of the incoming input digit, and $w_o$ the weight of the outgoing output digit. The initial state $e_C$ is $(0,\ 1/b_i,\ 1/b_o)$. The left algebra $(\triangleright_C)$ adds the product of the current input digit and its weight to the accumulator, and updates the input weight in preparation for the next input digit:

$$(\triangleright_C) : S_C \to \mathbb{N} \to S_C$$
$$(v,\ w_i,\ w_o) \triangleright_C d = (v + d \times w_i,\ w_i/b_i,\ w_o)$$

On the other hand, the coalgebra $g_C$ produces an output digit and updates the accumulator and the next output weight if the accumulator is not yet zero:

$$g_C : S_C \to \mathsf{Maybe}\,(\mathbb{N} \times S_C)$$
$$g_C\,(v,\ w_i,\ w_o) = \mathbf{if}\ v > 0\ \mathbf{then\ let}\ d = \lfloor v/w_o \rfloor; r = v - d \times w_o$$
$$\mathbf{in}\ just\,(d,\ (r,\ w_i,\ w_o/b_o))$$
$$\mathbf{else}\ nothing$$

For the example $0.625_{10} = 0.101_2$, the metamorphism first consumes the input digits using $(\triangleright_C)$,

$$(0,\ 0.1,\ 0.5) \overset{6}{\mapsto} (0.6,\ 0.01,\ 0.5) \overset{2}{\mapsto} (0.62,\ 0.001,\ 0.5) \overset{5}{\mapsto} (0.625,\ 0.0001,\ 0.5)$$

and then produces the output digits using $g_C$:

$$(0.625,\ 0.0001,\ 0.5) \overset{1}{\mapsto} (0.125,\ 0.0001,\ 0.25) \overset{0}{\mapsto} (0.125,\ 0.0001,\ 0.125)$$
$$\overset{1}{\mapsto} (0,\ 0.0001,\ 0.0625)\ \not\mapsto$$

## 3 | Specification of Metamorphisms in Types

In the rest of this paper we will develop (actually, reinvent) several *metamorphic algorithms*, which compute a metamorphism but do not take the form of a fold followed by an unfold. Rather than proving extrinsically that these algorithms satisfy their metamorphic specifications, we will encode metamorphic specifications intrinsically in types, such that any type-checked program is a correct metamorphic algorithm.

**Algebraic ornamentation**    The encoding is based on McBride's *algebraic ornamentation* [21]. Given a right algebra $(\triangleleft) : A \to S \to S$ and $e : S$, we can refine List $A$ into a family of types AlgList $A\,(\triangleleft)\,e : S \to$ Set (for 'algebraic lists') indexed by $S$ such that (conceptually) every list $as$ falls into the type AlgList $A\,(\triangleleft)\,e\,(foldr\,(\triangleleft)\,e\,as)$. The definition of AlgList is obtained by 'fusing' *foldr* into List:

**data** AlgList $(A : \mathsf{Set})\ \{S : \mathsf{Set}\}\ ((\triangleleft) : A \to S \to S)\ (e : S) : S \to \mathsf{Set}$ **where**
   [ ]  : AlgList $A\,(\triangleleft)\,e\,e$
   $\_::\_ : (a : A) \to \{s : S\} \to$ AlgList $A\,(\triangleleft)\,e\,s \to$ AlgList $A\,(\triangleleft)\,e\,(a \triangleleft s)$

The empty list is classified under the index $e = foldr\ (\lhd)\ e\ [\,]$. For the cons case, if a tail $as$ is classified under $s$, meaning that $foldr\ (\lhd)\ e\ as = s$, then the whole list $a :: as$ should be classified under $a \lhd s$ since $foldr\ (\lhd)\ e\ (a :: as) = a \lhd foldr\ (\lhd)\ e\ as = a \lhd s$. For example, we can obtain vectors, or length-indexed lists,

> **data** Vec $(A : \mathsf{Set}) : \mathbb{N} \to \mathsf{Set}$ **where**
> $\quad [\,] \quad : \mathsf{Vec}\ A\ 0$
> $\quad \_::\_ : (a : A) \to \{n : \mathbb{N}\} \to \mathsf{Vec}\ A\ n \to \mathsf{Vec}\ A\ (1 + n)$

simply by defining $\mathsf{Vec}\ A = \mathsf{AlgList}\ A\ (\lambda\ \_\ n \to 1 + n)\ 0$, since $length = foldr\ (\lambda\ \_\ n \to 1 + n)\ 0$.

**Coalgebraic ornamentation**   Dually, given a coalgebra $g : S \to \mathsf{Maybe}\ (B \times S)$, we can refine CoList $B$ into a family of types CoalgList $B\ g\ s : S \to \mathsf{Set}$ (for 'coalgebraic (co)lists') such that a colist falls into CoalgList $B\ g\ s$ if it is unfolded from $s$ using $g$. (Note that, extensionally, every CoalgList $B\ g\ s$ has exactly one inhabitant; intensionally there may be different ways to describe/compute that inhabitant, though.) Again the definition of CoalgList is obtained by fusing *unfoldr* into CoList:[5]

> **mutual**
> $\quad$ **record** CoalgList $(B : \mathsf{Set})\ \{S : \mathsf{Set}\}\ (g : S \to \mathsf{Maybe}\ (B \times S))\ (s : S) : \mathsf{Set}$ **where**
> $\qquad$ **coinductive**
> $\qquad$ **field**
> $\qquad\quad$ decon : CoalgListF $B\ g\ s$
> $\quad$ **data** CoalgListF $(B\ \{S\} : \mathsf{Set})\ (g : S \to \mathsf{Maybe}\ (B \times S))\ (s : S) : \mathsf{Set}$ **where**
> $\qquad \langle\_\rangle \qquad : g\ s \equiv \mathsf{nothing} \to \mathsf{CoalgListF}\ B\ g\ s$
> $\qquad \_::\langle\_\rangle\_ : (b : B) \to \{s' : S\} \to g\ s \equiv \mathsf{just}\ (b, s') \to$
> $\qquad\qquad\qquad \mathsf{CoalgList}\ B\ g\ s' \to \mathsf{CoalgListF}\ B\ g\ s$
>
> **open** CoalgList

Deconstructing a colist of type CoalgList $B\ g\ s$ can lead to two possible outcomes: the colist can be empty, in which case we also get an equality proof (enclosed in angle brackets) that $g\ s$ is nothing, or it can be non-empty, in which case we know that $g\ s$ produces the head element, and that the tail colist is unfolded from the next state $s'$ produced by $g\ s$.

Let $A, B, S : \mathsf{Set}$ throughout the rest of this paper (that is, think of the code in the rest of this paper as contained in a module with parameters $A, B, S : \mathsf{Set}$) — we will assume that $A$ is the type of input elements, $B$ the type of output elements, and $S$ the type of states. We will also consistently let $(\lhd) : A \to S \to S$ denote a right algebra, $(\rhd) : S \to A \to S$ a left algebra, $e : S$ an initial/empty state, and $g : S \to \mathsf{Maybe}\ (B \times S)$ a coalgebra.

---

[5] The operator $\_\equiv\_ : \{A : \mathsf{Set}\} \to A \to A \to \mathsf{Set}$ is the equality type constructor in the Agda Standard Library. An equality type $x \equiv y$ is inhabited by the refl constructor if $x$ has the same normal form as $y$, or uninhabited otherwise.

**Right metamorphisms**    Any program of type

$$\{s : S\} \to \mathsf{AlgList}\, A\, (\triangleleft)\, e\, s \to \mathsf{CoalgList}\, B\, g\, s$$

implements the right metamorphism *unfoldr g ∘ foldr (◁) e*, since the indexing enforces that the input list folds to *s*, from which the output colist is then unfolded.

**Left metamorphisms**    Conveniently, for left metamorphisms we do not need to define another variant of AlgList due to an old trick that expresses *foldl* in terms of *foldr*. Given a list *as* : List *A*, think of the work of *foldl* ($\triangleright$) *e as* as (i) partially applying *flip* ($\triangleright$) : *A* → *S* → *S* (where *flip f x y = f y x*) to every element of *as* to obtain state transformations of type *S* → *S*, (ii) composing the state transformations from left to right, and finally (iii) applying the resulting composite transformation to *e*. The left-to-right order appears only in step (ii), which, in fact, can also be performed from right to left since function composition is associative. Formally, we have

$$\mathit{foldl}\,(\triangleright)\, e\, as \;=\; \mathit{foldr}\,(\mathit{from\text{-}left\text{-}alg}\,(\triangleright))\, \mathit{id}\, as\, e$$

where

$$\mathit{from\text{-}left\text{-}alg} : \{A\, S : \mathsf{Set}\} \to (S \to A \to S) \to A \to (S \to S) \to (S \to S)$$
$$\mathit{from\text{-}left\text{-}alg}\,(\triangleright)\, a\, t \;=\; t \circ \mathit{flip}\,(\triangleright)\, a$$

and *id* : {*A* : Set} → *A* → *A* is the identity function. The type of left metamorphic algorithms can then be specified as

$$(s : S)\, \{h : S \to S\} \to \mathsf{AlgList}\, A\, (\mathit{from\text{-}left\text{-}alg}\,(\triangleright))\, \mathit{id}\, h \to \mathsf{CoalgList}\, B\, g\, (h\, s)$$

which says that if the initial state is *s* and the input list folds to a state transformation *h*, then the output colist should be unfolded from *h s*.

## 4    Metamorphisms, Definitionally

To warm up, let us start from the left metamorphic type and implement the most straightforward algorithm that strictly follows the definition of metamorphisms, **c**onsuming all inputs **b**efore **p**roducing outputs:

$$cbp : (s : S)\, \{h : S \to S\} \to \mathsf{AlgList}\, A\, (\mathit{from\text{-}left\text{-}alg}\,(\triangleright))\, \mathit{id}\, h \to \mathsf{CoalgList}\, B\, g\, (h\, s)$$
$$cbp\, s\, as\, {}_{\mathsf{AlgList}\, A\, (\mathit{from\text{-}left\text{-}alg}\,(\triangleright))\, \mathit{id}\, h} \;=\; \{\, \mathsf{CoalgList}\, B\, g\, (h\, s)\, \}_0$$

We will try to recreate in this paper what it feels like to program interactively with Agda.[6] During interaction, we can leave 'holes' in programs and fill or refine them, often with Agda's help.[7] Such a hole is called an *interaction point* or a *goal*, of which the green-shaded part above is an example. At goals, Agda can be instructed to provide various information and even perform some program synthesis. One most important piece of information for a goal is its expected type, which we always display in curly brackets. Goals are numbered so that they can be referred to in the text. At goals, we can also query the types of the variables in scope;[8] whenever the type of a variable needs to be displayed, we will annotate the variable with its type in yellow-shaded subscript (which is not part of the program text). In the program above, we annotate *as* with its type because the expected type at goal 0 refers to *h*, which is the index in the type of *as*.

---

[6] More precisely, it is the interactive development environment included in the Agda package that we interact with. This interactive development environment is implemented as an Emacs mode (and ported to some other editors as well), which explains the form of the commands that we will see in subsequent footnotes.

Now let us try to develop the program. We are trying to consume the input list first, so we pattern match on the argument *as* to see if there is anything to consume. We can instruct Agda to split the program into two clauses, listing all possible cases of *as*; goal 0 is gone as a result, and two new goals appear:

$$cbp : (s : S)\ \{h : S \rightarrow S\} \rightarrow \mathsf{AlgList}\ A\ (\textit{from-left-alg}\ (\triangleright))\ \textit{id}\ h \rightarrow \mathsf{CoalgList}\ B\ g\ (h\ s)$$
$$cbp\ s\ [\ ] = \{\ \mathsf{CoalgList}\ B\ g\ s\ \}_1$$
$$cbp\ s\ (a :: as_{\ \mathsf{AlgList}\ A\ (\textit{from-left-alg}\ (\triangleright))\ \textit{id}\ h}) = \{\ \mathsf{CoalgList}\ B\ g\ (h\ (s \triangleright a))\ \}_2$$

Note that the expected types of the two new goals have changed: at goal 1, for example, we see that the output colist should be unfolded directly from the initial state *s* since the input list is empty. By providing sufficient type information, Agda can keep track of such relationships for us! We continue to interact with and refine these two new goals.

**Consumption**  If there is something to consume, that is, the input list is non-empty, we go into goal 2. Here we should keep consuming the tail *as* but from a new state, so we refine goal 2 to goal 3 as follows:[9]

$$cbp : (s : S)\ \{h : S \rightarrow S\} \rightarrow \mathsf{AlgList}\ A\ (\textit{from-left-alg}\ (\triangleright))\ \textit{id}\ h \rightarrow \mathsf{CoalgList}\ B\ g\ (h\ s)$$
$$cbp\ s\ [\ ] = \{\ \mathsf{CoalgList}\ B\ g\ s\ \}_1$$
$$cbp\ s\ (a :: as) = cbp\ \{\ S\ \}_3\ as$$

What is this new state? It should be the one obtained by subsuming *a* into *s*, i.e., $s \triangleright a$. (Agda knows this too, in fact, and can fill the goal automatically.[10])

$$cbp : (s : S)\ \{h : S \rightarrow S\} \rightarrow \mathsf{AlgList}\ A\ (\textit{from-left-alg}\ (\triangleright))\ \textit{id}\ h \rightarrow \mathsf{CoalgList}\ B\ g\ (h\ s)$$
$$cbp\ s\ [\ ] = \{\ \mathsf{CoalgList}\ B\ g\ s\ \}_1$$
$$cbp\ s\ (a :: as) = cbp\ (s \triangleright a)\ as$$

---

[7] A hole can be created by putting in a question mark '?' and then instruct Agda to (re)load the program by giving the 'load' command (C-c C-l).

[8] The types of a goal and the variables in its context can be displayed by Agda upon receiving the 'goal type and context' command (C-c C-,).

[9] The refinement from goal 2 to goal 3 can be done by filling '*cbp* ? *as*' into goal 2 and giving the 'refine' command (C-c C-r) to Agda.

[10] This is done by giving the 'auto' command (C-c C-a), which invokes an external program Agsy [17] to search for a suitable inhabitant.

**Production**  If there is nothing more to consume, that is, the input list is empty, we go into goal 1, where we should produce the output colist; to specify the colist, we should say what will result if we deconstruct the colist. That is, we perform a copattern match:[11]

$$cbp : (s : S) \{h : S \to S\} \to \mathsf{AlgList}\, A\, (\mathit{from\text{-}left\text{-}alg}\, (\rhd))\, id\, h \to \mathsf{CoalgList}\, B\, g\, (h\, s)$$
$$decon\, (cbp\, s\, [\,]) = \{\, \mathsf{CoalgListF}\, B\, g\, s\, \}_4$$
$$cbp\, s\, (a :: as) = cbp\, (s \rhd a)\, as$$

The result of deconstruction depends on whether $g$ can produce anything from the current state $s$, so we pattern match on $g\, s$, splitting goal 4 into goals 5 and 6:

$$cbp : (s : S) \{h : S \to S\} \to \mathsf{AlgList}\, A\, (\mathit{from\text{-}left\text{-}alg}\, (\rhd))\, id\, h \to \mathsf{CoalgList}\, B\, g\, (h\, s)$$
$$decon\, (cbp\, s\, [\,]) \;\mathbf{with}\; g\, s$$
$$decon\, (cbp\, s\, [\,]) \mid nothing \quad = \{\, \mathsf{CoalgListF}\, B\, g\, s\, \}_5$$
$$decon\, (cbp\, s\, [\,]) \mid just\, (b,\, s') = \{\, \mathsf{CoalgListF}\, B\, g\, s\, \}_6$$
$$cbp\, s\, (a :: as) = cbp\, (s \rhd a)\, as$$

If $g\, s$ is nothing (goal 5), the output colist is empty; otherwise $g\, s$ is just $(b,\, s')$ for some $b$ and $s'$ (goal 6), in which case we use $b$ as the head and go on to produce the tail from $s'$. We therefore refine the two goals manually into

$$cbp : (s : S) \{h : S \to S\} \to \mathsf{AlgList}\, A\, (\mathit{from\text{-}left\text{-}alg}\, (\rhd))\, id\, h \to \mathsf{CoalgList}\, B\, g\, (h\, s)$$
$$decon\, (cbp\, s\, [\,]) \;\mathbf{with}\; g\, s$$
$$decon\, (cbp\, s\, [\,]) \mid nothing \quad = \langle\, \{\, g\, s \equiv nothing\, \}_7\, \rangle$$
$$decon\, (cbp\, s\, [\,]) \mid just\, (b,\, s') = b :: \langle\, \{\, g\, s \equiv just\, (b,\, s')\, \}_8\, \rangle\, cbp\, s'\, [\,]$$
$$cbp\, s\, (a :: as) = cbp\, (s \rhd a)\, as$$

We are now required to discharge equality proof obligations about $g\, s$, and the obligations exactly correspond to the results of the **with**-matching. This is precisely a situation in which the *inspect* idiom in the Agda Standard Library can help![12] With *inspect*, we can obtain an equality proof of the right type in each of the cases of the **with**-matching:

$$cbp : (s : S) \{h : S \to S\} \to \mathsf{AlgList}\, A\, (\mathit{from\text{-}left\text{-}alg}\, (\rhd))\, id\, h \to \mathsf{CoalgList}\, B\, g\, (h\, s)$$
$$decon\, (cbp\, s\, [\,]) \;\mathbf{with}\; g\, s \qquad\qquad\quad | \; inspect\, g\, s$$
$$decon\, (cbp\, s\, [\,]) \mid nothing \quad | [eq_{\,g\, s \equiv nothing}\quad\;] = \langle\, \{\, g\, s \equiv nothing\, \}_7\, \rangle$$
$$decon\, (cbp\, s\, [\,]) \mid just\, (b,\, s') \mid [eq_{\,g\, s \equiv just\, (b,\, s')}] = b :: \langle\, \{\, g\, s \equiv just\, (b,\, s')\, \}_8\, \rangle$$
$$\qquad\qquad\qquad\qquad\qquad\qquad\qquad\qquad\qquad\qquad cbp\, s'\, [\,]$$
$$cbp\, s\, (a :: as) = cbp\, (s \rhd a)\, as$$

Both goals can now be discharged with $eq$, and we arrive at a complete program, shown in figure 1. As explained in section 3, the correctness of this program is established by type-checking — no additional proofs are needed.

---

[11] This can be done by Agda if we give it the 'case split' command (C-c C-c) without specifying a variable.

[12] For more details about *inspect*, see https://agda.readthedocs.io/en/v2.6.1/language/with-abstraction.html#the-inspect-idiom.

**module** *ConsumingBeforeProducing*
$((\triangleright) : S \rightarrow A \rightarrow S)\,(g : S \rightarrow \text{Maybe}\,(B \times S))$
**where**

$cbp : (s : S)\,\{h : S \rightarrow S\} \rightarrow \text{AlgList}\,A\,(\textit{from-left-alg}\,(\triangleright))\,\textit{id}\,h \rightarrow \text{CoalgList}\,B\,g\,(h\,s)$
$decon\,(cbp\,s\,[\,])\,\textbf{with}\,g\,s \qquad |\,\textit{inspect}\,g\,s$
$decon\,(cbp\,s\,[\,])\,|\,\text{nothing} \quad |\,[\,eq\,] = \langle\,eq\,\rangle$
$decon\,(cbp\,s\,[\,])\,|\,\text{just}\,(b, s')\,|\,[\,eq\,] = b :: \langle\,eq\,\rangle\,cbp\,s'\,[\,]$
$cbp\,s\,(a :: as) = cbp\,(s \triangleright a)\,as$

■ **Figure 1**  Definitional implementation of metamorphisms

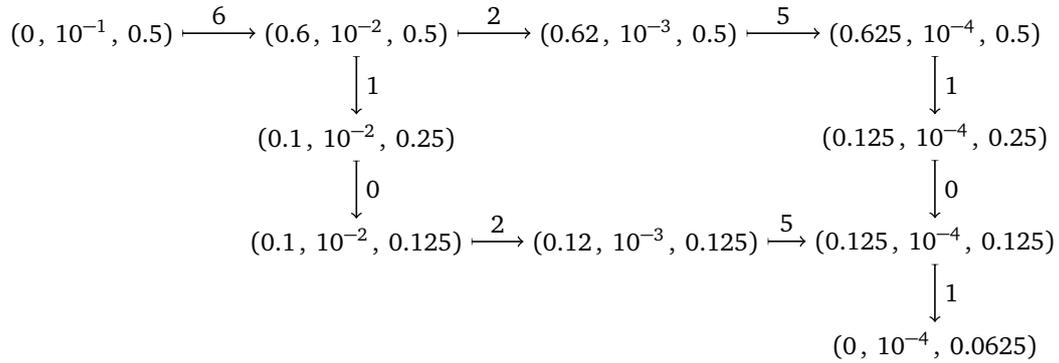

■ **Figure 2**  Streaming the conversion from $0.625_{10}$ to $0.101_2$

## 5  The Streaming Algorithm

As Gibbons [10] noted, metamorphisms in general cannot be automatically optimised in terms of time and space, but in some cases it is possible to compute a list metamorphism using a *streaming algorithm*, which can produce an initial segment of the output colist from an initial segment of the input list. For example (see figure 2), when converting $0.625_{10}$ to $0.101_2$, after consuming the first decimal digit 6 and reaching the state $(0.6,\,10^{-2},\,0.5)$, we can directly produce the first binary digit 1 because we know that the number will definitely be greater than 0.5. Streaming is not always possible, of course. Heapsort is a counterexample: no matter how many input elements have been consumed, it is always possible that a minimum element — which should be the first output element — has yet to appear, and thus we can never produce the first output element before we see the whole input list. There should be some condition under which we can stream metamorphisms, and we should be able to discover such condition if we program a streaming algorithm together with Agda, which knows what metamorphisms are and can provide us with semantic hints regarding what conditions need to be introduced to make the program a valid metamorphic algorithm.

We start from the same left metamorphic type:

$$stream : (s : S) \{h : S \to S\} \to$$
$$\text{AlgList } A \ (\textit{from-left-alg} \ (\rhd)) \ id \ h \to \text{CoalgList } B \ g \ (h \ s)$$
$$stream \ s \ as \ \boxed{\text{AlgList } A \ (\textit{from-left-alg} \ (\rhd)) \ id \ h} = \boxed{\{ \text{CoalgList } B \ g \ (h \ s) \}_0}$$

In contrast to *cbp* (section 4), this time we try to produce using *g* whenever possible, so our first step is to pattern match on *g s* (and we also introduce decon and *inspect*, which will be needed like in *cbp*):

$$stream : (s : S) \{h : S \to S\} \to$$
$$\text{AlgList } A \ (\textit{from-left-alg} \ (\rhd)) \ id \ h \to \text{CoalgList } B \ g \ (h \ s)$$
$$\text{decon } (stream \ s \ as \ \boxed{\text{AlgList } A \ (\textit{from-left-alg} \ (\rhd)) \ id \ h} ) \text{ with } g \ s \ | \ inspect \ g \ s$$
$$\text{decon } (stream \ s \ as) \ | \ \text{nothing} \quad | \ [eq] = \boxed{\{ \text{CoalgListF } B \ g \ (h \ s) \}_1}$$
$$\text{decon } (stream \ s \ as) \ | \ \text{just } (b, s') \ | \ [eq] = \boxed{\{ \text{CoalgListF } B \ g \ (h \ s) \}_2}$$

**Consumption**  For goal 1, we cannot produce anything since *g s* is nothing, but this does not mean that the output colist is empty — we may be able to produce something once we consume the input list and advance to a new state. We therefore pattern match on the input list, splitting goal 1 into goals 3 and 4:

$$stream : (s : S) \{h : S \to S\} \to$$
$$\text{AlgList } A \ (\textit{from-left-alg} \ (\rhd)) \ id \ h \to \text{CoalgList } B \ g \ (h \ s)$$
$$\text{decon } (stream \ s \ as) \text{ with } g \ s \quad | \ inspect \ g \ s$$
$$\text{decon } (stream \ s \ [\,]) \ | \ \text{nothing} \quad | \ [eq] = \boxed{\{ \text{CoalgListF } B \ g \ s \}_3}$$
$$\text{decon } (stream \ s \ (a :: as \ \boxed{\text{AlgList } A \ (\textit{from-left-alg} \ (\rhd)) \ id \ h} ))$$
$$\quad | \ \text{nothing} \ | \ [eq] = \boxed{\{ \text{CoalgListF } B \ g \ (h \ (s \rhd a)) \}_4}$$
$$\text{decon } (stream \ s \ as) \ | \ \text{just } (b, s') \ | \ [eq] = \boxed{\{ \text{CoalgListF } B \ g \ (h \ s) \}_2}$$

The two goals are similar to what we have seen in *cbp*. At goal 3, there is nothing more in the input list to consume, so we should end production and emit an empty colist, while for goal 4 we should advance to the new state $s \rhd a$ and set the tail *as* as the list to be consumed next:

$$stream : (s : S) \{h : S \to S\} \to$$
$$\text{AlgList } A \ (\textit{from-left-alg} \ (\rhd)) \ id \ h \to \text{CoalgList } B \ g \ (h \ s)$$
$$\text{decon } (stream \ s \ as \qquad) \text{ with } g \ s \quad | \ inspect \ g \ s$$
$$\text{decon } (stream \ s \ [\,] \qquad) \ | \ \text{nothing} \quad | \ [eq] = \langle \ eq \ \rangle$$
$$\text{decon } (stream \ s \ (a :: as)) \ | \ \text{nothing} \quad | \ [eq] = \text{decon } (stream \ (s \rhd a) \ as)$$
$$\text{decon } (stream \ s \ as \qquad) \ | \ \text{just } (b, s') \ | \ [eq] = \boxed{\{ \text{CoalgListF } B \ g \ (h \ s) \}_2}$$

**Production**  Goal 2 is the interesting case. Using *g*, from the current state *s* we can produce *b*, which we set as the head of the output colist, and advance to a new state $s'$, from which we produce the tail of the colist:

$$stream : (s : S) \{h : S \to S\} \to$$
$$\text{AlgList } A \ (\textit{from-left-alg} \ (\rhd)) \ id \ h \to \text{CoalgList } B \ g \ (h \ s)$$

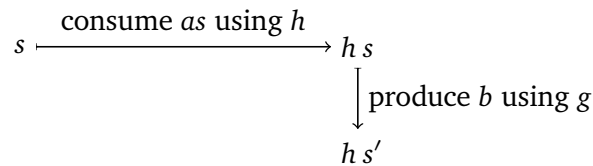

**The streaming condition**   Now Agda gives us a non-trivial proof obligation at goal 5 — what does it mean? The left-hand side $g(h\,s)$ is trying to produce using $g$ from the state $h\,s$, where $h$ is the state transformation resulting from consuming the entire input list $as$ (since $h$ is the index in the type of $as$), and the whole equality says that this has to produce a specific result. Drawing this as a state transition diagram:

$$s \xrightarrow{\text{consume } as \text{ using } h} h\,s \xrightarrow{\text{produce } b \text{ using } g} h\,s'$$

We already have in the context a similar-looking equality, namely $eq : g\,s \equiv \mathsf{just}\,(b\,,\,s')$, which we can superimpose on the diagram:

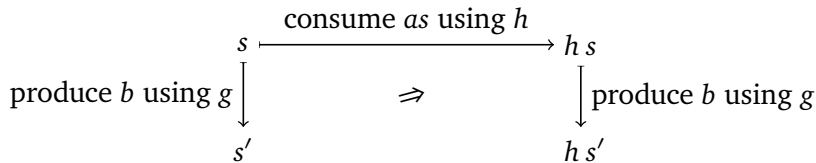

We also put in an implication arrow to indicate more explicitly that $g\,s \equiv \mathsf{just}\,(b\,,\,s')$ is a premise, from which we should derive $g\,(h\,s) \equiv \mathsf{just}\,(b\,,\,h\,s')$. Now it is tempting, and indeed easy, to complete the diagram:

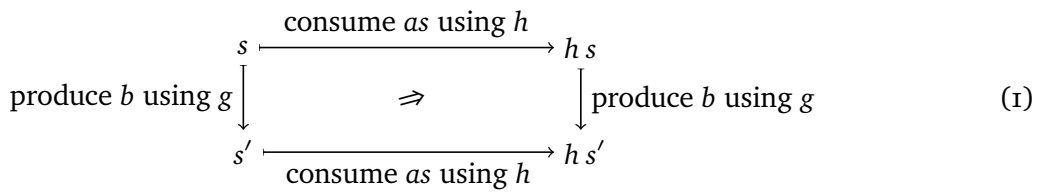   (1)

This is a kind of commutativity of production and consumption: From the initial state $s$, we can either

- apply $g$ to $s$ to produce $b$ and reach a new state $s'$, and then apply $h$ to consume the list and update the state to $h\,s'$, or
- apply $h$ to $s$ to consume the list and update the state to $h\,s$, and then apply $g$ to $h\,s$ to produce an element and reach a new state.

If the first route is possible, the second route should also be possible, and the outcomes should be the same — doing production using $g$ and consumption using $h$ in whichever order should emit the same element and reach the same final state. The two outcomes are not necessarily the same in general, though, and the commutativity should be formulated as a condition of the streaming algorithm.

But the above commutativity (1) of $g$ and $h$ — which is commutativity of one step of production (using $g$) and multiple steps of consumption (of the entire input list, using $h$) — may not be a good condition to impose. If we require instead that $g$ and ($\triangleright$) commute, this commutativity of single-step production and consumption will be easier for the algorithm user to verify:

$$
\begin{array}{ccc}
s & \xrightarrow{\;\text{consume } a \text{ using } (\triangleright)\;} & s \triangleright a \\
{\scriptstyle\text{produce } b \text{ using } g}\Big\downarrow & \Rrightarrow & \Big\downarrow{\scriptstyle\text{produce } b \text{ using } g} \\
s' & \xrightarrow[\;\text{consume } a \text{ using } (\triangleright)\;]{} & s' \triangleright a
\end{array}
\qquad (2)
$$

This is Gibbons's *streaming condition* [10, Definition 1]. In our development of *stream*, we need to assume that a proof of the streaming condition is available:

**constant** *streaming-condition* : $\{a : A\}\ \{b : B\}\ \{s\ s' : S\}\ \rightarrow$
$\qquad\qquad g\ s \equiv \mathsf{just}\ (b\ ,\ s') \rightarrow g\ (s \triangleright a) \equiv \mathsf{just}\ (b\ ,\ s' \triangleright a)$

We use a hypothetical **constant** keyword here to emphasise that *streaming-condition* is a constant made available to us and does not need to be defined. In the complete program, the functions defined in this section are collected in a module, and *streaming-condition* is made a parameter of this module.

**Wrapping up**    Back to goal 5, where we should prove the commutativity of $g$ and $h$. All it should take is a straightforward induction to extend the streaming condition along the axis of consumption. We know that we need a helper function *streaming-lemma* that performs induction on *as* and uses *eq* as a premise; Agda can help us to find the type,[13] and then we finish the rest, which is not hard:

*streaming-lemma* : $\{b : B\}\ \{s\ s' : S\}\ \{h : S \rightarrow S\}\ \rightarrow$
$\qquad\qquad$ AlgList $A$ (*from-left-alg* ($\triangleright$)) *id* $h$ $\rightarrow$
$\qquad\qquad g\ s \equiv \mathsf{just}\ (b\ ,\ s') \rightarrow g\ (h\ s) \equiv \mathsf{just}\ (b\ ,\ h\ s')$
*streaming-lemma* [ ]　　　　$eq = eq$
*streaming-lemma* ($a :: as$) $eq = $ *streaming-lemma as* (*streaming-condition eq*)

Agda then accepts *streaming-lemma as eq* as a type-correct term for goal 5, completing the definition of *stream*. The completed program is shown in figure 3.

**Streaming base conversion for fractions**    We have (re)discovered the streaming condition, but does it hold for the base conversion metamorphism given in section 2,

---

[13] More precisely, we can fill *streaming-lemma as eq* into goal 5 and give the 'helper type' command (C-c C-h), and Agda will generate a type for *streaming-lemma*. We have to manually remove some over-generalisations and unnecessary definition expansions though.

**module** *Streaming*
  $((\rhd) : S \rightarrow A \rightarrow S) \, (g : S \rightarrow \mathsf{Maybe}\,(B \times S))$
  $(streaming\text{-}condition : \{a : A\}\,\{b : B\}\,\{s\,s' : S\} \rightarrow$
                      $g\,s \equiv \mathsf{just}\,(b,\,s') \rightarrow g\,(s \rhd a) \equiv \mathsf{just}\,(b,\,s' \rhd a))$
  **where**
  $streaming\text{-}lemma : \{b : B\}\,\{s\,s' : S\}\,\{h : S \rightarrow S\} \rightarrow \mathsf{AlgList}\,A\,(from\text{-}left\text{-}alg\,(\rhd))\,id\,h \rightarrow$
                        $g\,s \equiv \mathsf{just}\,(b,\,s') \rightarrow g\,(h\,s) \equiv \mathsf{just}\,(b,\,h\,s')$
  $streaming\text{-}lemma\,[\,]\qquad eq = eq$
  $streaming\text{-}lemma\,(a :: as)\,eq = streaming\text{-}lemma\,as\,(streaming\text{-}condition\,eq)$

  $stream : (s : S)\,\{h : S \rightarrow S\} \rightarrow \mathsf{AlgList}\,A\,(from\text{-}left\text{-}alg\,(\rhd))\,id\,h \rightarrow \mathsf{CoalgList}\,B\,g\,(h\,s)$
  $decon\,(stream\,s\,as\qquad)$ **with** $g\,s\qquad\quad |$ *inspect* $g\,s$
  $decon\,(stream\,s\,[\,]\qquad)\,|\,\mathsf{nothing}\quad |\,[eq] = \langle\,eq\,\rangle$
  $decon\,(stream\,s\,(a :: as))\,|\,\mathsf{nothing}\quad |\,[eq] = decon\,(stream\,(s \rhd a)\,as)$
  $decon\,(stream\,s\,as\qquad)\,|\,\mathsf{just}\,(b,\,s')\,|\,[eq] = b :: \langle\,streaming\text{-}lemma\,as\,eq\,\rangle\,stream\,s'\,as$

■ **Figure 3** The streaming algorithm

namely *unfoldr* $g_C \circ foldl\,(\rhd_C)\,e_C$? Actually, no. The problem is that $g_C$ can be too eager to produce an output digit. In $0.625_{10} = 0.101_2$, for example, after consuming the first decimal digit 6, we can safely use $g_C$ to produce the first two binary digits 1 and 0, reaching the state $(0.1,\ 0.01,\ 0.125)$. From this state, $g_C$ will produce a third binary digit 0, but this can be wrong if there are more input digits to consume — indeed, in our example the next input digit is 5, and the accumulator will go up to $0.1 + 5 \times 0.01 = 0.15$, exceeding the next output weight $0.125$, and hence the next output digit should be 1 instead. To allow streaming, we should make $g_C$ more conservative, producing an output digit only when the accumulator will not go up too much to change the produced output digit whatever the unconsumed input digits might be. We therefore revise $g_C$ to check an extra condition (underlined below) before producing:

$$g'_C\,(v,\,w_i,\,w_o) = \textbf{let}\,d = \lfloor v/w_o \rfloor; r = v - d \times w_o$$
$$\textbf{in}\,\ \textbf{if}\,v > 0 \wedge \underline{r + b_i \times w_i \leqslant w_o}\,\textbf{then}\,\mathsf{just}\,(d,\,(r,\,w_i,\,w_o/b_o))$$
$$\textbf{else}\,\ \mathsf{nothing}$$

In this extra condition, $r$ is the updated accumulator after producing an output digit, and $b_i \times w_i$ is the supremum value attainable by the unconsumed input digits. If the sum $r + b_i \times w_i$ exceeds $w_o$, the output digit may have to be increased, in which case we should not produce the digit just yet. After this revision, the streaming condition holds for $g'_C$ and $(\rhd_C)$.

Once all the input digits have been consumed, however, $g'_C$ can be too conservative and does not produce output digits even when the accumulator is not yet zero. This is another story though, which is told by Gibbons [10, Section 4.4].

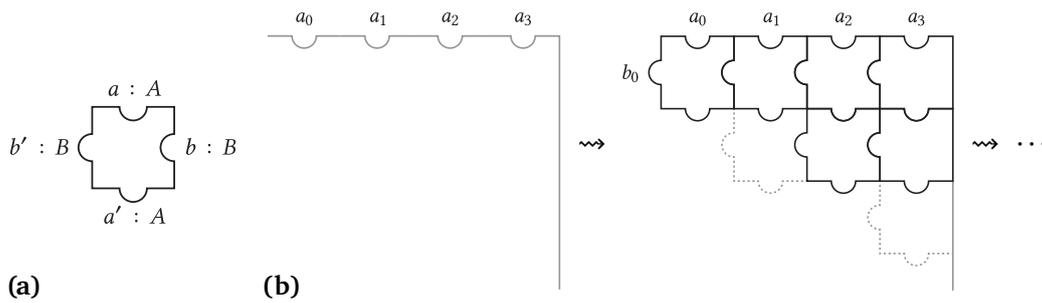

■ **Figure 4**  The jigsaw model

## 6 The Jigsaw Model

Let us now turn to right metamorphisms. Recall that the right metamorphic type is

$$\{s : S\} \ \to \ \mathsf{AlgList}\, A\, (\lhd)\, e\, s \ \to \ \mathsf{CoalgList}\, B\, g\, s$$

which, unlike the left metamorphic type, does not have an initial state as one of its arguments. The only state appearing in the type is the implicit argument $s : S$, which is the intermediate state reached after consuming the entire input list, and it is unreasonble to assume that this intermediate state is also given at the start of a metamorphic computation — for example, when computing the heapsort metamorphism, we would not expect a heap containing all the elements of the input list to be given to an algorithm as input. This suggests that $s$ plays a role only in the type-level specification, and we should avoid using $s$ in the actual computation — that is, $s$ should be computationally irrelevant, and could be somehow erased. Correspondingly, the indices and proofs in AlgList and CoalgList could all be erased eventually, turning a program with the right metamorphic type into one that maps plain lists to colists. Does this mean that we can bypass computation with states and just work with list elements to compute a metamorphism? Surprisingly, Nakano has such a computation model [23], in which it is possible to compute a metamorphism without using the states mentioned in its specification! (By contrast, in *cbp* (section 4) and *stream* (section 5), we can hope to erase the indices and proofs in AlgList and CoalgList but not the input state, which is used relevantly in the computation.)

### 6.1 Computation in the Jigsaw Model

In Nakano's model [23], a computation transforms a List $A$ to a CoList $B$, and to program its behaviour, we need to provide a suitable function *piece* : $A \times B \ \to \ B \times A$. Nakano neatly visualises his model as a jigsaw puzzle. The *piece* function can be thought of as describing a set of jigsaw pieces (which are not to be rotated or flipped) illustrated in figure 4(a). In each piece, the horizontal edges are associated with a value of type $A$, and the vertical edges with a value of type $B$. Two pieces fit together exactly when the values on their adjacent edges coincide. Moreover, the values on the top and right edges should determine those on the left and bottom edges, and the *piece* function records the mappings for all the pieces — the piece above, for example, corresponds to the mapping *piece* $(a , b) = (b', a')$.

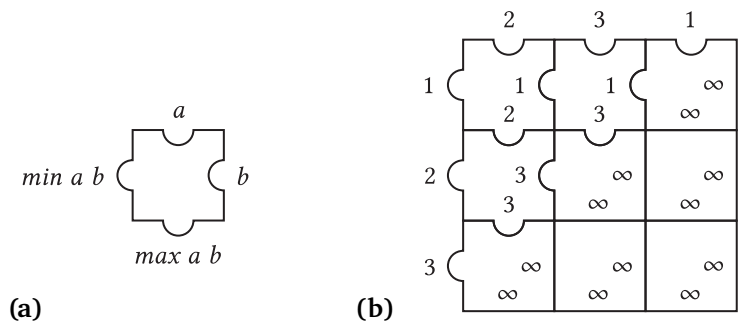

**■ Figure 5** Heapsort in the jigsaw model

Figure 4(b) is an illustration of an ongoing computation: Given an input list, say $a_0 :: a_1 :: a_2 :: a_3 :: [\,]$, we start from an empty board with its top boundary initialised to the input elements and its right boundary to some special 'straight' value. Then we put in pieces to fill the board: Whenever a top edge and an adjacent right edge is known, we consult the *piece* function to find the unique fitting piece and put it in. Initially the only place where we can put a piece is the top-right corner, but as we put in more pieces, the number of choices will increase — in the board on the right, for example, we can choose one of the two dashed places to put the next piece in. Eventually we will reach the left boundary, and the values on the left boundary are the output elements. Although we can put in the pieces in a nondeterministic order and even in parallel, the final board configuration is determined by the initial boundary, and thus the output elements are produced deterministically.

**Heapsort in the jigsaw model** For a concrete example, the heapsort metamorphism can be computed in the jigsaw model with *piece* $(a, b) = (min\ a\ b, max\ a\ b)$, as illustrated in figure 5(a). The final board configuration after sorting the list $2 :: 3 :: 1 :: [\,]$ is shown in figure 5(b). Nakano remarks that computing heapsort in the jigsaw model transforms heapsort into 'a form of parallel bubble sort', which looks very different from the original metamorphic computation — in particular, heaps are nowhere to be seen.

In general, how is the jigsaw model related to metamorphisms, and under what conditions does the jigsaw model compute metamorphisms? Again, we will figure out the answers by programming jigsaw computations with metamorphic types in Agda.

## 6.2 The Infinite Case

We will only look at a simpler case where the output colist is always infinite; that is, the coalgebra used in the metamorphic type is just $\circ\ g^\infty$ where $g^\infty : S \to B \times S$ — for heapsort, $g^\infty$ is an adapted version of *popMin* such that popping from the empty heap returns $\infty$ and the empty heap itself, so the output colist is the sorted input list followed by an infinite number of $\infty$'s. The general case (where the output colist may be finite) is more complicated and not included in the main text, but a solution can be found in appendix A.

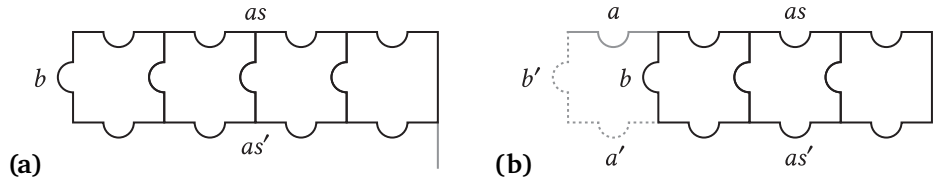

■ **Figure 6** Steps of horizontal placement

### 6.2.1 Horizontal Placement

We start by giving the metamorphic type signature (where the subscript IH is short for 'infinite' and 'horizontal'):

$jigsaw_{IH} : \{s : S\} \to \mathsf{AlgList}\,A\,(\lhd)\,e\,s \to \mathsf{CoalgList}\,B\,(\mathsf{just} \circ g^{\infty})\,s$

$jigsaw_{IH}\,as_{\boxed{\mathsf{AlgList}\,A\,(\lhd)\,e\,s}} = \boxed{\{\,\mathsf{CoalgList}\,B\,(\mathsf{just} \circ g^{\infty})\,s\,\}_0}$

One possible placement strategy is to place one row of jigsaw pieces at a time. As illustrated in figure 6(a), placing a row is equivalent to transforming an input list *as* into a new one *as′* and also a vertical edge *b*. We therefore introduce the following function $fill_{IH}$ for filling a row:[14]

$fill_{IH} : \{s : S\} \to \mathsf{AlgList}\,A\,(\lhd)\,e\,s \to B \times \Sigma[\,t \in S\,]\,\mathsf{AlgList}\,A\,(\lhd)\,e\,t$

$fill_{IH}\,as = \boxed{\{\,B \times \Sigma[\,t \in S\,]\,\mathsf{AlgList}\,A\,(\lhd)\,e\,t\,\}_1}$

We do not know (or cannot easily specify) the index $t$ in the type of the new AlgList, so the index is simply existentially quantified. The job of $jigsaw_{IH}$, then, is to call $fill_{IH}$ repeatedly to cover the board. We revise goal 0 into

$jigsaw_{IH} : \{s : S\} \to \mathsf{AlgList}\,A\,(\lhd)\,e\,s \to \mathsf{CoalgList}\,B\,(\mathsf{just} \circ g^{\infty})\,s$

$\mathsf{decon}\,(jigsaw_{IH}\,as_{\boxed{\mathsf{AlgList}\,A\,(\lhd)\,e\,s}})\,\mathbf{with}\,fill_{IH}\,as$

$\mathsf{decon}\,(jigsaw_{IH}\,as) \mid b,\,t,\,as' = b ::\langle\,\boxed{\{\,\mathsf{just}\,(g^{\infty}\,s) \equiv \mathsf{just}\,(b,\,t)\,\}_2}\,\rangle\,jigsaw_{IH}\,as'$

Goal 2 demands an equality linking $s$ and $t$, which are the input and output indices of $fill_{IH}$; this suggests that $fill_{IH}$ is responsible for not only computing $t$ but also establishing the relationship between $t$ and $s$. We therefore add the equality to the result type of $fill_{IH}$, and discharge goal 2 with the equality proof that will be produced by $fill_{IH}$:

$fill_{IH} : \{s : S\} \to \mathsf{AlgList}\,A\,(\lhd)\,e\,s \to$
$\qquad \Sigma[\,b \in B\,]\,\Sigma[\,t \in S\,]\,\mathsf{AlgList}\,A\,(\lhd)\,e\,t \times (g^{\infty}\,s \equiv (b,\,t))$

$fill_{IH}\,as = \boxed{\{\,\Sigma[\,b \in B\,]\,\Sigma[\,t \in S\,]\,\mathsf{AlgList}\,A\,(\lhd)\,e\,t \times (g^{\infty}\,s \equiv (b,\,t))\,\}_1}$

$jigsaw_{IH} : \{s : S\} \to \mathsf{AlgList}\,A\,(\lhd)\,e\,s \to \mathsf{CoalgList}\,B\,(\mathsf{just} \circ g^{\infty})\,s$

$\mathsf{decon}\,(jigsaw_{IH}\,as)\,\mathbf{with}\,fill_{IH}\,as$

$\mathsf{decon}\,(jigsaw_{IH}\,as) \mid b,\,\_,\,as',\,eq = b ::\langle\,cong\,\mathsf{just}\,eq\,\rangle\,jigsaw_{IH}\,as'$

$\quad - cong\,\mathsf{just}\,\{X : \mathsf{Set}\}\,\{x\,x' : X\} \to x \equiv x' \to \mathsf{just}\,x \equiv \mathsf{just}\,x'$

---

[14] $\Sigma[\,x \in X\,]\,P\,x$ is the dependent pair type from the Agda Standard Library.

**Computational irrelevance**  From goal 2, there seems to be another way forward: the equality says that the output vertical edge $b$ and the index $t$ in the type of $as'$ are determined by $g^\infty s$, so $jigsaw_\text{IH}$ could have computed $b$ and $t$ directly! However, recall that the characteristic of the jigsaw model is that computation proceeds by converting input list elements directly into output colist elements without involving the states appearing in the metamorphic specification. In our setting, this means that states only appear in the function types, not the function bodies, so having $jigsaw_\text{IH}$ invoke $g\ s$ would deviate from the jigsaw model. Instead, $jigsaw_\text{IH}$ invokes $fill_\text{IH}$, which will only use *piece* to compute $b$. (It would probably be better if we declared the argument $s$ in the metamorphic type as irrelevant to enforce the fact that $s$ does not participate in the computation; this irrelevance declaration would then need to be propagated to related parts in AlgList and CoalgList, though, which is a complication we want to avoid.)

**Filling a row**  Let us get back to $fill_\text{IH}$ (goal 1). The process of filling a row follows the structure of the input list, so overall it is an induction, of which the first step is a case analysis:

$fill_\text{IH} : \{s : S\} \to \text{AlgList}\,A\,(\lhd)\,e\,s \to$
$\qquad \Sigma[\,b \in B\,]\,\Sigma[\,t \in S\,]\,\text{AlgList}\,A\,(\lhd)\,e\,t \times (g^\infty\,s \equiv (b,\,t))$
$fill_\text{IH}\,[\,] = \{\Sigma[\,b \in B\,]\,\Sigma[\,t \in S\,]\,\text{AlgList}\,A\,(\lhd)\,e\,t \times (g^\infty\,e \equiv (b,\,t))\}_3$
$fill_\text{IH}\,(a :: as\ _{\text{AlgList}\,(\lhd)\,e\,s}) =$
$\qquad \{\Sigma[\,b \in B\,]\,\Sigma[\,t \in S\,]\,\text{AlgList}\,A\,(\lhd)\,e\,t \times (g^\infty\,(a \lhd s) \equiv (b,\,t))\}_4$

If the input list is empty (goal 3), we return the rightmost 'straight' edge. We therefore assume that a **constant** *straight* : $B$ is available, and fill it into goal 3:

$fill_\text{IH} : \{s : S\} \to \text{AlgList}\,A\,(\lhd)\,e\,s \to$
$\qquad \Sigma[\,b \in B\,]\,\Sigma[\,t \in S\,]\,\text{AlgList}\,A\,(\lhd)\,e\,t \times (g^\infty\,s \equiv (b,\,t))$
$fill_\text{IH}\,[\,] = straight,\ \{\Sigma[\,t \in S\,]\,\text{AlgList}\,A\,(\lhd)\,e\,t \times (g^\infty\,e \equiv (straight,\,t))\}_5$
$fill_\text{IH}\,(a :: as\ _{\text{AlgList}\,(\lhd)\,e\,s}) =$
$\qquad \{\Sigma[\,b \in B\,]\,\Sigma[\,t \in S\,]\,\text{AlgList}\,A\,(\lhd)\,e\,t \times (g^\infty\,(a \lhd s) \equiv (b,\,t))\}_4$

At goal 5, we should now give the new list (along with the index in its type), which we know should have the same length as the old list, so in this case it is easy to see that the new list should be empty as well (and we leave the index in the type of the new list for Agda to infer by giving an underscore):

$fill_\text{IH} : \{s : S\} \to \text{AlgList}\,A\,(\lhd)\,e\,s \to$
$\qquad \Sigma[\,b \in B\,]\,\Sigma[\,t \in S\,]\,\text{AlgList}\,A\,(\lhd)\,e\,t \times (g^\infty\,s \equiv (b,\,t))$
$fill_\text{IH}\,[\,] = straight,\ \_,\ [\,],\ \{g^\infty\,e \equiv (straight,\,e)\}_6$
$fill_\text{IH}\,(a :: as\ _{\text{AlgList}\,(\lhd)\,e\,s}) =$
$\qquad \{\Sigma[\,b \in B\,]\,\Sigma[\,t \in S\,]\,\text{AlgList}\,A\,(\lhd)\,e\,t \times (g^\infty\,(a \lhd s) \equiv (b,\,t))\}_4$

Here we arrive at another proof obligation (goal 6), which says that from the initial state $e$ the coalgebra $g^\infty$ should produce *straight* and leave the state unchanged. This seems a reasonable property to add as a condition of the algorithm: in heapsort, for example, $e$ is the empty heap and *straight* is $\infty$, and popping from the empty heap, as we mentioned, can be defined to return $\infty$ and the empty heap itself. We therefore add an additional **constant** *straight-production* : $g^\infty e \equiv (straight, e)$, which discharges goal 6.

The interesting case is when the input list is non-empty (goal 4). We start with an inductive call to $fill_{IH}$ itself:

$$fill_{IH} : \{s : S\} \rightarrow \mathsf{AlgList}\, A\, (\lhd)\, e\, s \rightarrow$$
$$\Sigma[\, b \in B\, ]\, \Sigma[\, t \in S\, ]\, \mathsf{AlgList}\, A\, (\lhd)\, e\, t \times (g^\infty s \equiv (b, t))$$
$$fill_{IH}\, [\,] = straight,\, \_,\, [\,],\, straight\text{-}production$$
$$fill_{IH}\, (a :: as\ \boxed{\mathsf{AlgList}\,(\lhd)\,e\,s})\, \mathbf{with}\, fill_{IH}\, as$$
$$fill_{IH}\, (a :: as\qquad\quad)\ |\ b,\, s',\, as',\, eq =$$
$$\boxed{\{\,\Sigma[\, b \in B\, ]\, \Sigma[\, t \in S\, ]\, \mathsf{AlgList}\, A\, (\lhd)\, e\, t \times (g^\infty\, (a \lhd s) \equiv (b, t))\,\}_7}$$

As illustrated in figure 6(b), the inductive call places the jigsaw pieces below the tail $as$, yielding a vertical edge $b$ and a list $as'$ of horizontal edges below $as$. We should complete the row by placing the last jigsaw piece with $a$ and $b$ as input, and use the output edges $(b', a') = piece\, (a, b)$ in the right places:

$$fill_{IH} : \{s : S\} \rightarrow \mathsf{AlgList}\, A\, (\lhd)\, e\, s \rightarrow$$
$$\Sigma[\, b \in B\, ]\, \Sigma[\, t \in S\, ]\, \mathsf{AlgList}\, A\, (\lhd)\, e\, t \times (g^\infty s \equiv (b, t))$$
$$fill_{IH}\, [\,] = straight,\, \_,\, [\,],\, straight\text{-}production$$
$$fill_{IH}\, (a :: as\ \boxed{\mathsf{AlgList}\,(\lhd)\,e\,s})\, \mathbf{with}\, fill_{IH}\, as$$
$$fill_{IH}\, (a :: as\qquad\quad)\ |\ (b,\, s',\, as',\, eq_{\boxed{g^\infty s \equiv (b, s')}}) =$$
$$\mathbf{let}\, (b',\, a') = piece\, (a, b)\, \mathbf{in}\, b',\, \_,\, a' :: as',\ \boxed{\{\,g^\infty\, (a \lhd s) \equiv (b', a' \lhd s')\,\}_8}$$

**The jigsaw condition**   Here we see a familiar pattern: goal 8 demands an equality about producing from a state after consumption, and in the context we have an equality $eq$ about producing from a state before consumption. Following what we did in section 5, a commutative state transition diagram can be drawn:

$$
\begin{array}{ccc}
a \lhd s & \xleftarrow{\text{consume } a \text{ using } (\lhd)} & s \\
{\scriptstyle\text{produce } b' \text{ using } g^\infty}\downarrow & \Leftarrow & \downarrow{\scriptstyle\text{produce } b \text{ using } g^\infty} \\
a' \lhd s' & \xleftarrow{\text{consume } a' \text{ using } (\lhd)} & s'
\end{array}
\qquad (3)
$$

where $(b', a') = piece\, (a, b)$. Diagram (3) is again a kind of commutativity of production and consumption, but unlike the streaming condition (2) in section 5, the elements produced and consumed can change after *swapping* the order of production and consumption. Given any top and right edges $a$ and $b$, the *piece* function should be able to find the left and bottom edges $b'$ and $a'$ to complete the commutative diagram. This requirement constitutes a specification for *piece*, and, inspired by Nakano [23] (but not following him strictly), we call it the *jigsaw condition*:

$$\mathbf{constant}\, jigsaw\text{-}condition_1 : \{a : A\}\, \{b : B\}\, \{s\, s' : S\} \rightarrow g^\infty s \equiv (b, s') \rightarrow$$
$$\mathbf{let}\, (b',\, a') = piece\, (a, b)\, \mathbf{in}\, g^\infty\, (a \lhd s) \equiv (b', a' \lhd s')$$

Adding *jigsaw-condition*$_1$ as the final assumption, we can fill *jigsaw-condition*$_1$ *eq* into goal 8 and complete the program, which is shown in figure 7.

**Metamorphisms and the jigsaw model**   We can now see a connection between metamorphic computations and the jigsaw model. Definitionally, a metamorphism folds the input list to a state, and then produces the output elements while updating the state. In the jigsaw model, and with the horizontal placement strategy, rather than folding the input list to a 'compressed' state, we use the whole list as an 'uncompressed' state, and ensure that the production process using uncompressed states simulates the definitional one using compressed states. The type of *fill*$_{\text{IH}}$ makes this clear: The produced element *b* is exactly the one that would have been produced from the compressed state *s* obtained by folding the old list. Then, on the compressed side, the state *s* is updated to *t*; correspondingly, on the uncompressed side, the old list is updated to a new list that folds to *t*. The jigsaw condition ensures that this relationship between compressed and uncompressed states can be maintained by placing rows of jigsaw pieces. In the case of heapsort, this can be understood as directly using lists to represent heaps — observe in figure 5(b) that placing a row of jigsaw pieces is equivalent to extracting the minimum element as the leftmost edge and leaving the rest of the elements on the bottom edges of the row, and the whole computation proceeds like selection sort.

**Deriving the *piece* function for heapsort using the jigsaw condition**   For the heapsort metamorphism, consuming an element is pushing it into a heap, and producing an element is popping a minimum element from a heap. In diagram (3), producing *b* on the right means that *b* is a minimum element in the heap *s*, and *s'* is the rest of the heap. If *a* is pushed into *s*, popping from the updated heap $a \lhd s$ will either still produce *b* if $a > b$, or produce *a* if $a \leqslant b$, so *b'* should be *min a b*. Afterwards, the final heap $a' \lhd s'$ should still contain the other element that was not popped out, i.e., *max a b*, and can be obtained by pushing *max a b* into *s'*, so *a'* should be *max a b*.

### 6.2.2 Vertical Placement

There is another obvious placement strategy: the vertical one, where we place one column of jigsaw pieces at a time. Programming the horizontal placement strategy led us to an understanding of the relationship between metamorphisms and the jigsaw model, and it is likely that programming the vertical placement strategy will lead us to a different perspective. Starting from exactly the same type signature as *jigsaw*$_{\text{IH}}$ (except the function name, where the subscript IV is short for 'infinite' and 'vertical'):

$jigsaw_{\text{IV}} : \{s : S\} \rightarrow \text{AlgList } A (\lhd) e s \rightarrow \text{CoalgList } B (\text{just} \circ g^{\infty}) s$
$jigsaw_{\text{IV}} \ as \ _{\boxed{\text{AlgList } A (\lhd) e s}} \ = \ \{ \text{CoalgList } B (\text{just} \circ g^{\infty}) s \}_0$

With the vertical placement strategy, we place columns of jigsaw pieces following the structure of the input list *as*, so we proceed with a case analysis on *as*:

$jigsaw_{\text{IV}} : \{s : S\} \rightarrow \text{AlgList } A (\lhd) e s \rightarrow \text{CoalgList } B (\text{just} \circ g^{\infty}) s$
$jigsaw_{\text{IV}} \ [\,] \ = \ \{ \text{CoalgList } B (\text{just} \circ g^{\infty}) e \}_1$
$jigsaw_{\text{IV}} \ (a :: as \ _{\boxed{\text{AlgList } A (\lhd) e s}}) \ = \ \{ \text{CoalgList } B (\text{just} \circ g^{\infty}) (a \lhd s) \}_2$

**module** *Jigsaw-Infinite-Horizontal*
  $((\lhd) : A \to S \to S)\,(e : S)\,(g^{\infty} : S \to B \times S)$
  $(piece : A \times B \to B \times A)$
  $(straight : B)\,(straight\text{-}production : g^{\infty}\,e \equiv (straight,\,e))$
  $(jigsaw\text{-}condition_1 : \{a : A\}\,\{b : B\}\,\{s\,s' : S\} \to$
               $g^{\infty}\,s \equiv (b,\,s') \to \mathbf{let}\,(b',\,a') = piece\,(a,\,b)\,\mathbf{in}\,g^{\infty}\,(a \lhd s) \equiv (b',\,a' \lhd s'))$

**where**

$fill_{\mathrm{IH}} : \{s : S\} \to \mathsf{AlgList}\,A\,(\lhd)\,e\,s \to \Sigma[\,b \in B\,]\,\Sigma[\,t \in S\,]\,\mathsf{AlgList}\,A\,(\lhd)\,e\,t \times (g^{\infty}\,s \equiv (b,\,t))$
$fill_{\mathrm{IH}}\,[\,] = straight\,,\,\_\,,\,[\,]\,,\,straight\text{-}production$
$fill_{\mathrm{IH}}\,(a :: as)\,\mathbf{with}\,fill_{\mathrm{IH}}\,as$
$fill_{\mathrm{IH}}\,(a :: as)\mid b\,,\,\_\,,\,as'\,,\,eq = \mathbf{let}\,(b',\,a') = piece\,(a,\,b)$
                           $\mathbf{in}\,\,b'\,,\,\_\,,\,a' :: as'\,,\,jigsaw\text{-}condition_1\,eq$

$jigsaw_{\mathrm{IH}} : \{s : S\} \to \mathsf{AlgList}\,A\,(\lhd)\,e\,s \to \mathsf{CoalgList}\,B\,(just \circ g^{\infty})\,s$
$\mathsf{decon}\,(jigsaw_{\mathrm{IV}}\,as)\,\mathbf{with}\,fill_{\mathrm{IH}}\,as$
$\mathsf{decon}\,(jigsaw_{\mathrm{IV}}\,as)\mid b\,,\,\_\,,\,as'\,,\,eq = b :: \langle\,cong\,just\,eq\,\rangle\,jigsaw_{\mathrm{IV}}\,as'$

■ **Figure 7** Metamorphisms in the infinite jigsaw model with the horizontal placement strategy

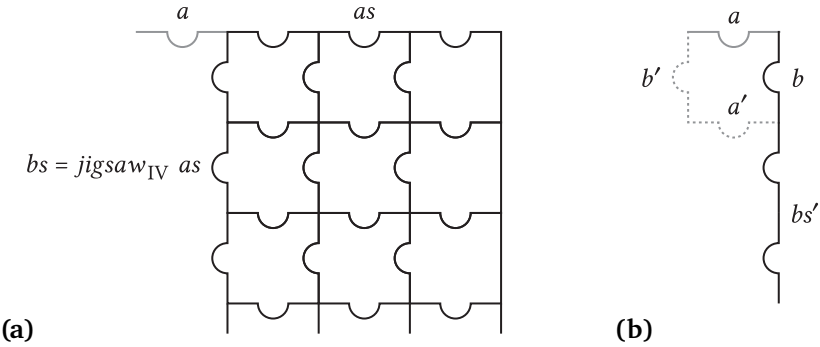

$bs = jigsaw_{\mathrm{IV}}\,as$

**(a)**                                **(b)**

■ **Figure 8** Steps of vertical placement

If the input list is empty (goal 1), we should produce a colist of *straight* egdes:

$jigsaw_{\mathrm{IV}} : \{s : S\} \to \mathsf{AlgList}\,A\,(\lhd)\,e\,s \to \mathsf{CoalgList}\,B\,(just \circ g^{\infty})\,s$
$\mathsf{decon}\,(jigsaw_{\mathrm{IV}}\,[\,]) = straight :: \langle\,\boxed{\{\,just\,(g^{\infty}\,e) \equiv just\,(straight,\,e)\,\}_3}\,\rangle\,jigsaw_{\mathrm{IV}}\,[\,]$
$jigsaw_{\mathrm{IV}}\,(a :: as\,{}_{\boxed{\mathsf{AlgList}\,A\,(\lhd)\,e\,s}}) = \boxed{\{\,\mathsf{CoalgList}\,B\,(just \circ g^{\infty})\,(a \lhd s)\,\}_2}$

The proof obligation (goal 3) is discharged with *cong just straight-production*, where both *straight* and *straight-production* are constants we introduced when programming the horizontal placement strategy (section 6.2.1). The inductive case (goal 2) is depicted in figure 8(a): We place all the columns below the tail *as* by an inductive call $jigsaw_{\mathrm{IV}}\,as$, which gives us a colist of vertical edges. To the left of this colist, we should place the last column below the head element *a*; again we introduce a helper function $fill_{\mathrm{IV}}$ that takes *a* and the colist $jigsaw_{\mathrm{IV}}\,as$ as input and produces the leftmost colist:

$jigsaw_{\mathrm{IV}} : \{s : S\} \to \mathsf{AlgList}\,A\,(\lhd)\,e\,s \to \mathsf{CoalgList}\,B\,(just \circ g^{\infty})\,s$
$\mathsf{decon}\,(jigsaw_{\mathrm{IV}}\,[\,]) = straight :: \langle\,cong\,just\,straight\text{-}production\,\rangle\,jigsaw_{\mathrm{IV}}\,[\,]$
$jigsaw_{\mathrm{IV}}\,(a :: as) = fill_{\mathrm{IV}}\,a\,(jigsaw_{\mathrm{IV}}\,as)$

**Filling a column**  Agda again can help us to find a suitable type of $fill_{IV}$:

$fill_{IV} : \{s : S\} (a : A) \rightarrow \text{CoalgList } B (\text{just} \circ g^{\infty}) s \rightarrow \text{CoalgList } B (\text{just} \circ g^{\infty}) (a \lhd s)$
$fill_{IV} \; a \; bs \; {}_{\text{CoalgList } B (\text{just} \circ g^{\infty}) s} \;=\; \{ \text{CoalgList } B (\text{just} \circ g^{\infty}) (a \lhd s) \}_4$

Here we should deconstruct $bs$ so that we can invoke $piece$ on $a$ and the first element of $bs$:

$fill_{IV} : \{s : S\} (a : A) \rightarrow \text{CoalgList } B (\text{just} \circ g^{\infty}) s \rightarrow \text{CoalgList } B (\text{just} \circ g^{\infty}) (a \lhd s)$
$\text{decon } (fill_{IV} \; a \; bs \; {}_{\text{CoalgList } B (\text{just} \circ g^{\infty}) s}) \textbf{ with } \text{decon } bs$
$\text{decon } (fill_{IV} \; a \; bs) \mid \langle \; eq \; {}_{\text{just } (g^{\infty} s) \equiv \text{nothing}} \; \rangle \;=\; \{ \text{CoalgListF } B (\text{just} \circ g^{\infty}) (a \lhd s) \}_5$
$\text{decon } (fill_{IV} \; a \; bs) \mid b ::\langle \; eq \; {}_{\text{just } (g^{\infty} s) \equiv \text{just } (b, \, s')} \; \rangle \; bs' \; {}_{\text{CoalgList } B (\text{just} \circ g^{\infty}) s'} \;=\;$
$\qquad\qquad\qquad\qquad\qquad \{ \text{CoalgListF } B (\text{just} \circ g^{\infty}) (a \lhd s) \}_6$

For goal 5, since the coalgebra $\text{just} \circ g^{\infty}$ in the type of $bs$ never returns nothing, it is impossible for $bs$ to be empty:[15]

$fill_{IV} : \{s : S\} (a : A) \rightarrow \text{CoalgList } B (\text{just} \circ g^{\infty}) s \rightarrow \text{CoalgList } B (\text{just} \circ g^{\infty}) (a \lhd s)$
$\text{decon } (fill_{IV} \; a \; bs \; {}_{\text{CoalgList } B (\text{just} \circ g^{\infty}) s}) \textbf{ with } \text{decon } bs$
$\text{decon } (fill_{IV} \; a \; bs) \mid \langle \, () \, \rangle$
$\text{decon } (fill_{IV} \; a \; bs) \mid b ::\langle \; eq \; {}_{\text{just } (g^{\infty} s) \equiv \text{just } (b, \, s')} \; \rangle \; bs' \; {}_{\text{CoalgList } B (\text{just} \circ g^{\infty}) s'} \;=\;$
$\qquad\qquad\qquad\qquad\qquad \{ \text{CoalgListF } B (\text{just} \circ g^{\infty}) (a \lhd s) \}_6$

The real work is done at goal 6, where $bs$ is deconstructed into its head $b$ and tail $bs'$ — see figure 8(b). We invoke the $piece$ function to transform $a$ and $b$ into $b'$ and $a'$; the head of the output colist is then $b'$, and the tail is coinductively computed from $a'$ and $bs'$:

$fill_{IV} : \{s : S\} (a : A) \rightarrow \text{CoalgList } B (\text{just} \circ g^{\infty}) s \rightarrow \text{CoalgList } B (\text{just} \circ g^{\infty}) (a \lhd s)$
$\text{decon } (fill_{IV} \; a \; bs \; {}_{\text{CoalgList } B (\text{just} \circ g^{\infty}) s}) \textbf{ with } \text{decon } bs$
$\text{decon } (fill_{IV} \; a \; bs) \mid \langle \, () \, \rangle$
$\text{decon } (fill_{IV} \; a \; bs) \mid b ::\langle \; eq \; {}_{\text{just } (g^{\infty} s) \equiv \text{just } (b, \, s')} \; \rangle \; bs' \; {}_{\text{CoalgList } B (\text{just} \circ g^{\infty}) s'} \;=\;$
$\qquad \textbf{let } (b', \, a') = piece \, (a, \, b)$
$\qquad \textbf{in } \; b' ::\langle \, \{ \text{just } (g^{\infty} (a \lhd s)) \equiv \text{just } (b', \, a' \lhd s') \}_7 \, \rangle \; fill_{IV} \; a' \; bs'$

The remaining proof obligation is exactly the jigsaw condition (3) modulo the harmless occurrences of just, so we arrive at a complete program, shown in figure 9, which uses the same set of conditions as the horizontal placement strategy.

**Metamorphisms and the jigsaw model revisited**  Now we can see that the vertical placement strategy indeed corresponds to another way of thinking about metamorphic computations in the jigsaw model. In contrast to streaming metamorphisms (section 5), where we need to be cautious about producing an element because once an element

---

[15] We can convince Agda that this case is impossible by matching $eq$ with the absurd pattern (), saying that $eq$ cannot possibly exist (and Agda accepts this because a just-value can never be equal to nothing).

```
module Jigsaw-Infinite-Vertical
   ((◁) : A → S → S) (e : S) (g∞ : S → B × S)
   (piece : A × B → B × A)
   (jigsaw-condition₁ : {a : A} {b : B} {s s′ : S} →
                          g∞ s ≡ (b , s′) → let (b′ , a′) = piece (a , b) in g∞ (a ◁ s) ≡ (b′ , a′ ◁ s′))
   (straight : B) (straight-production : g∞ e ≡ (straight , e))
   where
   fill_IV : {s : S} (a : A) → CoalgList B (just ∘ g∞) s → CoalgList B (just ∘ g∞) (a ◁ s)
   decon (fill_IV a bs) with decon bs
   decon (fill_IV a bs) | ⟨ () ⟩
   decon (fill_IV a bs) | b ::⟨ eq ⟩ bs′ =
      let (b′ , a′) = piece (a , b)
      in  b′ ::⟨ cong just (jigsaw-condition₁ (cong-from-just eq)) ⟩ fill_IV a′ bs′
         – cong-from-just : {X : Set} {x x′ : X} → just x ≡ just x′ → x ≡ x′
   jigsaw_IV : {s : S} → AlgList A (◁) e s → CoalgList B (just ∘ g∞) s
   decon (jigsaw_IV [ ]) = straight ::⟨ cong just straight-production ⟩ jigsaw_IV [ ]
   jigsaw_IV (a :: as) = fill_IV a (jigsaw_IV as)
```

■ **Figure 9**   Metamorphisms in the infinite jigsaw model with the vertical placement strategy is produced we can no longer change it, computing metamorphisms in the jigsaw model with the vertical placement strategy is like having an entire output colist right from the start and then updating it:

- initially we start with a colist of *straight* edges, which is unfolded from the empty state *e*;
- inductively, if we have a colist unfolded from some state *s*, and an input element *a* comes in, we place a column of jigsaw pieces to update the colist, and the result — due to the jigsaw condition — is a colist unfolded from the new state $a ◁ s$;
- finally, after all elements of the input list *as* are consumed, we get a colist unfolded from *foldr* (◁) *e as*.

Notably, the inductive step is faithfully described by the type of $fill_{IV}$ (which was generated by Agda). In the case of heapsort, this can be understood as putting (finite) elements into a colist with only ∞'s and keeping the elements in order — observe in figure 5(b) that placing a column of jigsaw pieces is equivalent to inserting the element on the top edge into the (sorted) colist on the right edges of the column, and the whole computation proceeds like insertion sort.

## 7  Discussion

We end our experiment at this point, but there are more to explore. It was slightly mysterious that we chose to implement the streaming algorithm with the left metamorphic type and the jigsaw model with the right metamorphic type — is this pairing necessary or coincidental? It turns out that the jigsaw model also works for left metamorphisms, and we even get a slightly more general algorithm if we start from the left metamorphic type — this is not fleshed out here, but the same type-driven methodology works too. (One solution is included in the supplementary code.) Streaming, on the other hand, seems to be inherently associated with left metamorphisms. One way to explain this might be as follows: During streaming, we consume an initial segment of an input list, pause to do some production, and then resume consumption of the rest of the list. The natural way to do this is to consume the list from the left, examining and removing head elements and keeping the tail for the resumption of production. That is, we are really treating the input list as a (finite) colist. This suggests that a streaming algorithm in general should be a transformation from colists to colists, both possibly infinite — that is, it is in general a 'stream processor' [9]. We might also consider using the jigsaw model to implement stream processors, but the situation can be more complicated: if we think of computation in the jigsaw model as updating the output colist (section 6.2.2), in general the output colist can change forever so that we are never sure whether any of the output elements has stabilised. More fundamentally, the current metamorphic specification — a fold followed by an unfold — is no longer adequate: since the input colist can be infinite, we have to replace the fold with some infinite consumption process, meaning that in general we can no longer reach an intermediate state and switch to production. We have to go back to the basics and think about how we might specify the behaviour of stream processors, and tell a new story from there.

This paper is mostly a faithful report of the actual developments of the programs from their specifications (as types). There are only minor deviations from the actual developments for streamlining the presentation, and apart from the general ideas of streaming and the jigsaw model, few hints were taken from the original papers by Gibbons [10] and Nakano [23]. For example, from goal 0 to goals 1 and 2 in section 5, bringing in decon and *inspect* was really a decision that could only be made later (at goal 3). And a step that might have been influenced by Gibbons was the leap from diagram (1) to diagram (2), i.e., Gibbons's streaming condition, but Agda did take us to diagram (1), which was probably close enough to the streaming condition. We certainly did not rely on any of the proofs in the original papers, simply because we did not have to construct most of them. For example, if we compare our development of the streaming algorithm (section 5) with that of Bird and Gibbons's [4] (who gave the original formulation and proof of the streaming theorem), we will see that their Lemma 29 turns into our *streaming-lemma* and their Theorem 30 goes implicitly into the typing of *stream* and no longer needs special attention. Notably, our formal treatment of the jigsaw model is independent and evidently different from Nakano's — the whole section 6 can be seen as an attempt to understand the relationship between metamorphisms and the jigsaw model without looking into Nakano's proofs.

So what does the experiment tell us about interactive type-driven development? Fundamentally, our development is just the familiar mathematical activity of formulating conditions of theorems by trying to prove the theorems and finding out what is missing. Agda does make the process unusually smooth, though. In addition to the automation provided by Agda, we speculate that the smoothness is due to what we might call 'proof locality', by which we mean that proofs appear near where they matter. As a result, the programmer gains better 'situation awareness' of what a program means while constructing the program. There are other desirable forms of situation awareness, for example the ability to execute incomplete programs offered by Hazel [27], and in general it would be interesting to develop more mechanisms to improve the programmer's situation awareness.

The experiment is far from decisive when it comes to determining the effectiveness of the methodology of interactive type-driven development, though. As mentioned in the beginning, we deliberately chose a problem that was not too complicated to make the experiment more likely to succeed. If we aim for a more sophisticated problem, for example, 'metaphorisms' [26], which are metamorphisms whose result is optimised in some way, it will probably require more advanced techniques like that of Ko and Gibbons's [15], where a type is first transformed into one that is closer to program structure. More generally, to respond to the question posed in section 1 'how well does interactive type-driven development scale?', we can consider scalability in the direction of algorithmic sophistication, and experiment with more algorithmic problems, departing from the well-understood examples such as bounds/dimensions and typed embedded languages, and expanding the applicability of the methodology.

## Acknowledgements

The author thanks Shin-Cheng Mu for a discussion during ICFP 2017 in Oxford, Jeremy Gibbons for discussions in Oxford and Tokyo and for reading and commenting on a draft of this paper, and the anonymous reviewers for ICFP 2018 and the *Programming* journal for their comments and suggestions. This work originated from the author's DPhil work [12], which was supported by a University of Oxford Clarendon Scholarship and the UK Engineering and Physical Sciences Research Council (EPSRC) project *Reusability and Dependent Types*. Before taking up the current post, the author was employed at the National Institute of Informatics in Tokyo, Japan, where the work continued with the support of the Japan Society for the Promotion of Science (JSPS) Grant-in-Aid for Scientific Research (A) No. 25240009 and (S) No. 17H06099. The author is currently supported by the Ministry of Science and Technology (MOST), Taiwan under grant No. 109-2222-E-001-002-MY3.

## A  Metamorphisms in the Jigsaw Model: The General Case

Here we tackle the general case of jigsaw metamorphisms, where the produced colist can be finite. This is the same setting as Nakano's [23], and will allow us to compare our derived conditions with his, although the details can get slightly complicated. The metamorphic type we use is exactly the one we saw in section 3:

$$jigsaw : \{s : S\} \rightarrow \mathsf{AlgList}\,A\,(\lhd)\,e\,s \rightarrow \mathsf{CoalgList}\,B\,g\,s$$
$$jigsaw\ as\ _{\mathsf{AlgList}\,A\,(\lhd)\,e\,s}\ =\ \{\,\mathsf{CoalgList}\,B\,g\,s\,\}_0$$

We use the vertical placement strategy, so the overall structure will be similar to $jigsaw_{\mathrm{IV}}$ (section 6.2.2).

The beginning is somewhat routine. Starting from a case analysis:

$$jigsaw : \{s : S\} \rightarrow \mathsf{AlgList}\,A\,(\lhd)\,e\,s \rightarrow \mathsf{CoalgList}\,B\,g\,s$$
$$jigsaw\ [\,]\ =\ \{\,\mathsf{CoalgList}\,B\,g\,e\,\}_1$$
$$jigsaw\ (a :: as\ _{\mathsf{AlgList}\,A\,(\lhd)\,e\,s}\ )\ =\ \{\,\mathsf{CoalgList}\,B\,g\,(a \lhd s)\,\}_2$$

At goal 1, it should suffice to produce an empty colist:

$$jigsaw : \{s : S\} \rightarrow \mathsf{AlgList}\,A\,(\lhd)\,e\,s \rightarrow \mathsf{CoalgList}\,B\,g\,s$$
$$decon\ (jigsaw\ [\,])\ =\ \langle\ \{\,g\,e \equiv \mathsf{nothing}\,\}_3\ \rangle$$
$$jigsaw\ (a :: as\ _{\mathsf{AlgList}\,A\,(\lhd)\,e\,s}\ )\ =\ \{\,\mathsf{CoalgList}\,B\,g\,(a \lhd s)\,\}_2$$

To do so we need $g\,e \equiv \mathsf{nothing}$, which is a reasonable assumption — for heapsort, for example, $e$ is the empty heap, on which *popMin* computes to nothing. We therefore introduce a **constant** *nothing-from-e* : $g\,e \equiv \mathsf{nothing}$ and use it to discharge goal 3:

$$jigsaw : \{s : S\} \rightarrow \mathsf{AlgList}\,A\,(\lhd)\,e\,s \rightarrow \mathsf{CoalgList}\,B\,g\,s$$
$$decon\ (jigsaw\ [\,])\ =\ \langle\ \textit{nothing-from-e}\ \rangle$$
$$jigsaw\ (a :: as\ _{\mathsf{AlgList}\,A\,(\lhd)\,e\,s}\ )\ =\ \{\,\mathsf{CoalgList}\,B\,g\,(a \lhd s)\,\}_2$$

For goal 2, we proceed in exactly the same way as we dealt with the corresponding case of $jigsaw_{\mathrm{IV}}$:

$$jigsaw : \{s : S\} \rightarrow \mathsf{AlgList}\,A\,(\lhd)\,e\,s \rightarrow \mathsf{CoalgList}\,B\,g\,s$$
$$decon\ (jigsaw\ [\,])\ =\ \langle\ \textit{nothing-from-e}\ \rangle$$
$$jigsaw\ (a :: as)\ =\ \textit{fill}\ a\ (jigsaw\ as)$$

where the type and the top-level structure of the helper function *fill* is also exactly the same as $fill_{\mathrm{IV}}$:

$$fill : \{s : S\}\,(a : A) \rightarrow \mathsf{CoalgList}\,B\,g\,s \rightarrow \mathsf{CoalgList}\,B\,g\,(a \lhd s)$$
$$decon\ (\textit{fill}\ a\ bs\ _{\mathsf{CoalgList}\,B\,g\,s}\ )\ \mathbf{with}\ decon\ bs$$
$$decon\ (\textit{fill}\ a\ bs)\ \mid\ \langle\ eq\ _{g\,s \equiv \mathsf{nothing}}\ \rangle\ =\ \{\,\mathsf{CoalgListF}\,B\,g\,(a \lhd s)\,\}_4$$
$$decon\ (\textit{fill}\ a\ bs)\ \mid\ b ::\langle\ eq\ _{g\,s \equiv \mathsf{just}\,(b,\,s')}\ \rangle\ bs'\ _{\mathsf{CoalgList}\,B\,g\,s'}\ =$$
$$\{\,\mathsf{CoalgListF}\,B\,g\,(a \lhd s)\,\}_5$$

Things get more interesting from this point.

**Filling a column**   Let us work on the familiar case first, namely goal 5. If we do the same thing as the corresponding case of $fill_{IV}$,

$fill : \{s : S\}\ (a : A) \rightarrow \mathsf{CoalgList}\ B\ g\ s \rightarrow \mathsf{CoalgList}\ B\ g\ (a \lhd s)$
$\mathsf{decon}\ (fill\ a\ bs\ \boxed{\mathsf{CoalgList}\ B\ g\ s}\ )\ \textbf{with}\ \mathsf{decon}\ bs$
$\mathsf{decon}\ (fill\ a\ bs)\ \mid\ \langle\ eq\ \boxed{g\ s \equiv \mathsf{nothing}}\ \rangle\ =\ \boxed{\{\ \mathsf{CoalgListF}\ B\ g\ (a \lhd s)\}_4}$
$\mathsf{decon}\ (fill\ a\ bs)\ \mid\ b ::\langle\ eq\ \boxed{g\ s \equiv \mathsf{just}\ (b,\ s')}\ \rangle\ bs'\ \boxed{\mathsf{CoalgList}\ B\ g\ s'}\ =$
$\quad \textbf{let}\ (b',\ a') = piece\ (a,\ b)\ \textbf{in}\ b' ::\langle\ \boxed{\{\ g\ (a \lhd s) \equiv \mathsf{just}\ (b',\ a' \lhd s')\}_6}\ \rangle\ fill\ a'\ bs'$

we will see that the condition we need is depicted in the same way as diagram (3) for the infinite jigsaw condition. Formally it is slightly different though, because we need to wrap the results of $g$ in the just constructor:

$$\{a : A\}\ \{b : B\}\ \{s\ s' : S\} \rightarrow$$
$$g\ s \equiv \mathsf{just}\ (b, s') \rightarrow \textbf{let}\ (b', a') = piece\ (a, b)\ \textbf{in}\ g\ (a \lhd s) \equiv \mathsf{just}\ (b', a' \lhd s')) \qquad (4)$$

We will come back to this condition and close goal 6 later.

Goal 4, unlike the corresponding case of $fill_{IV}$, is no longer an impossible case. We might be tempted to produce an empty colist here,

$fill : \{s : S\}\ (a : A) \rightarrow \mathsf{CoalgList}\ B\ g\ s \rightarrow \mathsf{CoalgList}\ B\ g\ (a \lhd s)$
$\mathsf{decon}\ (fill\ a\ bs\ \boxed{\mathsf{CoalgList}\ B\ g\ s}\ )\ \textbf{with}\ \mathsf{decon}\ bs$
$\mathsf{decon}\ (fill\ a\ bs)\ \mid\ \langle\ eq\ \boxed{g\ s \equiv \mathsf{nothing}}\ \rangle\ =\ \langle\ \boxed{\{\ g\ (a \lhd s) \equiv \mathsf{nothing}\}_7}\ \rangle$
$\mathsf{decon}\ (fill\ a\ bs)\ \mid\ b ::\langle\ eq\ \boxed{g\ s \equiv \mathsf{just}\ (b,\ s')}\ \rangle\ bs'\ \boxed{\mathsf{CoalgList}\ B\ g\ s'}\ =$
$\quad \textbf{let}\ (b',\ a') = piece\ (a,\ b)\ \textbf{in}\ b' ::\langle\ \boxed{\{\ g\ (a \lhd s) \equiv \mathsf{just}\ (b',\ a' \lhd s')\}_6}\ \rangle\ fill\ a'\ bs'$

but the proof obligation indicates that this is not a right choice. Let us call a state $s$ 'empty' exactly when $g\ s \equiv \mathsf{nothing}$. The proof obligation says that if a state $s$ is empty then the next state $a \lhd s$ should also be empty, but this does not hold in general — for heapsort, pushing a finite element into a heap always results in a non-empty heap, constituting a counterexample. On the other hand, it is conceivable that we can make some elements satisfy this property — for example, it is reasonable to define the *push* operation on heaps such that pushing $\infty$ into an empty heap keeps the heap empty — so producing an empty colist is not always wrong.

**Flat elements**   The above reasoning suggests that we should do a case analysis on $a$ to determine whether to produce an empty or non-empty colist at goal 4. Let us call an element 'flat' exactly when subsuming it into an empty state results in another empty state. We should be given a decision procedure $flat^?$ that can be used to identify flat elements:

$\textbf{constant}\ flat^? : (a : A) \rightarrow$
$$(\{s : S\} \rightarrow g\ s \equiv \mathsf{nothing} \rightarrow g\ (a \lhd s) \equiv \mathsf{nothing}) \uplus \boxed{\{\ \mathsf{Set}\}_8}$$

Traditionally $flat^?$ would return a boolean, but using booleans in dependently typed programming is almost always a 'code smell' since their meaning — for example, whether the input satisfies a certain property or not — will almost always need to be explained to the type-checker later; instead, it is more convenient to make the decision procedure directly return a proof or a refutation of the property. In the case of $flat^?$, its type directly says that an element of $A$ is flat or otherwise. This 'otherwise' at goal 8 also requires some thought. We could fill in the negation of the 'flat' property, but it may turn out that we need something stronger. Unable to decide now, let us leave goal 8 open for the moment, and come back when we have more information.

Abandoning goal 7, we roll back to goal 4 and refine it into goals 9 and 10:

$fill : \{s : S\} (a : A) \to \text{CoalgList } B\, g\, s \to \text{CoalgList } B\, g\, (a \lhd s)$
$\text{decon } (fill\, a\, bs\ \boxed{\text{CoalgList } B\, g\, s}\ ) \textbf{ with } \text{decon } bs$
$\text{decon } (fill\, a\, bs) \mid \langle\, eq\, \boxed{g\, s \equiv \text{nothing}}\, \rangle \textbf{ with } flat^?\, a$
$\text{decon } (fill\, a\, bs) \mid \langle\, eq\, \rangle \mid \text{inj}_1\, flat \quad\ = \langle\, \{\, g\, (a \lhd s) \equiv \text{nothing}\, \}_9\, \rangle$
$\text{decon } (fill\, a\, bs) \mid \langle\, eq\, \rangle \mid \text{inj}_2\, not\text{-}flat\ = \boxed{\{\text{CoalgListF } B\, g\, (a \lhd s)\}_{10}}$
$\text{decon } (fill\, a\, bs) \mid b :: \langle\, eq\, \boxed{g\, s \equiv \text{just}\, (b,\, s')}\, \rangle\, bs'\ \boxed{\text{CoalgList } B\, g\, s'}\ =$
$\quad \textbf{let } (b',\, a') = piece\, (a,\, b) \textbf{ in } b' :: \langle\, \{\, g\, (a \lhd s) \equiv \text{just}\, (b',\, a' \lhd s')\, \}_6\, \rangle\, fill\, a'\, bs'$

At goal 9, we know that $a$ is flat, so it is fine to produce an empty colist; the proof obligation is easily discharged with $flat\, eq$, where $flat$ is the proof given by $flat^?$ affirming that $a$ is flat.

For goal 10, we want to invoke $piece$ and produce a non-empty colist. However, the input colist is empty, so we do not have a vertical input edge for $piece$. The situation is not entirely clear here, but let us make some choices first and see if they make sense later. Without an input vertical edge, let us again introduce a **constant** $straight : B$, which solves the problem about using $piece$. Also, in the coinductive call that generates the tail, we use $bs$ (the only colist available in the context) as the second argument:

$fill : \{s : S\} (a : A) \to \text{CoalgList } B\, g\, s \to \text{CoalgList } B\, g\, (a \lhd s)$
$\text{decon } (fill\, a\, bs\ \boxed{\text{CoalgList } B\, g\, s}\ ) \textbf{ with } \text{decon } bs$
$\text{decon } (fill\, a\, bs) \mid \langle\, eq\, \boxed{g\, s \equiv \text{nothing}}\, \rangle \textbf{ with } flat^?\, a$
$\text{decon } (fill\, a\, bs) \mid \langle\, eq\, \rangle \mid \text{inj}_1\, flat \quad\ = \langle\, flat\, eq\, \rangle$
$\text{decon } (fill\, a\, bs) \mid \langle\, eq\, \rangle \mid \text{inj}_2\, not\text{-}flat\ =$
$\quad \textbf{let } (b',\, a') = piece\, (a,\, straight) \textbf{ in } b' :: \langle\, \{\, g\, (a \lhd s) \equiv \text{just}\, (b',\, a' \lhd s)\, \}_{11}\, \rangle\, fill\, a'\, bs$
$\text{decon } (fill\, a\, bs) \mid b :: \langle\, eq\, \boxed{g\, s \equiv \text{just}\, (b,\, s')}\, \rangle\, bs'\ \boxed{\text{CoalgList } B\, g\, s'}\ =$
$\quad \textbf{let } (b',\, a') = piece\, (a,\, b) \textbf{ in } b' :: \langle\, \{\, g\, (a \lhd s) \equiv \text{just}\, (b',\, a' \lhd s')\, \}_6\, \rangle\, fill\, a'\, bs'$

**The jigsaw condition**  Now let us examine whether our choices are sensible. The expected type at goal 11 can be depicted as

$$
\begin{array}{ccc}
& \xleftarrow{\text{consume } a \text{ using } f} & \\
a \lhd s & & s \\
\text{produce } b' \text{ using } g \downarrow & \quad \Leftarrow & \vdots \ (\text{produce } straight \text{ using } g) \qquad (5) \\
a' \lhd s & & s \\
& \xrightarrow{\text{consume } a' \text{ using } f} &
\end{array}
$$

The dashed transition on the right is not a real state transition — we know that $s$ is an empty state since in the context we have $eq : g\, s \equiv \text{nothing}$. Completing the above diagram (5) with the dashed transition allows us to compare it with diagram (3) for the infinite jigsaw condition, and the key to the comparison is to link the notions of empty states in the infinite case and the general (possibly finite) case. In the infinite case, we have a condition *straight-production* : $g^\infty\, e \equiv (straight\,,\, e)$ saying that the *straight* edge is produced from the empty state $e$, which remains unchanged after production. We could have defined empty states in the infinite case to be the states $s$ such that $g^\infty\, s \equiv (straight\,,\, s)$ (although this was not necessary). Now, the general (possibly finite) case can be thought of as an optimisation of the infinite case. We stop producing *straight* edges from empty states — that is, we modify the coalgebra to return nothing from empty states — because these *straight* edges provide no information: if we omit, and only omit, the production of these *straight* edges, then whenever a vertical input edge is missing we know that it can only be *straight*. However, the modification to the coalgebra destroys the production transitions from empty states in the infinite jigsaw condition. What remains is condition (4), where cases involving empty states and *straight* edges as depicted by diagram (5) above are left out.

One thing we can do is to merge diagram (5) back into condition (4) by relaxing the latter's premise:

> **constant** *jigsaw-condition* :
> $\qquad \{a : A\}\,\{b : B\}\,\{s\,s' : S\}\ \to$
> $\qquad (g\,s \equiv \mathsf{just}\,(b\,,\,s'))\ \uplus\ ((g\,s \equiv \mathsf{nothing})\ \times$
> $\qquad\qquad\qquad\qquad\qquad (g\,(a \lhd s) \not\equiv \mathsf{nothing})\times(b \equiv straight)\times(s' \equiv s))\ \to$
> $\qquad \mathbf{let}\ (b'\,,\,a') = piece\,(a\,,\,b)\ \mathbf{in}\ g\,(a \lhd s) \equiv \mathsf{just}\,(b'\,,\,a' \lhd s')$

Note that we include $g\,(a \lhd s) \not\equiv \mathsf{nothing}$ in the new part of the premise to rule out the case where $a$ is flat. A proof of this type should come from $flat^?$, so at goal 8 we make $flat^?$ return a proof of $\{s : S\}\ \to\ g\,(a \lhd s) \not\equiv \mathsf{nothing}$ when the input element $a$ is not flat. (Note that this type is not just the negation of flatness.) Finally, having *jigsaw-condition* in the context is informative enough for Agda to discharge both goals 6 and 11 automatically, and we arrive at a complete program (figure 10).

**Comparison with Nakano's jigsaw condition**    How do our conditions compare with Nakano's [23, Definition 5.1]? Ours seem to be weaker, but this is probably because our algorithm is not as sophisticated as it could be. Nakano imposes three conditions, which he refers to collectively as the jigsaw condition: the first one is exactly our *nothing-from-e*, the second one is related to flat elements but more complicated than our corresponding formulation, and the third one, though requiring some decoding, is almost our *jigsaw-condition*. Comparing Nakano's third condition with our *jigsaw-condition* reveals that there was one possibility that we did not consider: at goal 5 we went ahead and produced a non-empty colist, but producing an empty colist was also a possibility. Our current *jigsaw* algorithm places columns of non-decreasing lengths from right to left like **module** *Jigsaw-General*
  $((\triangleleft) : A \to S \to S)\, (e : S)\, (g : S \to \mathsf{Maybe}\,(B \times S))$
  $(\textit{piece} : A \times B \to B \times A)$
  $(\textit{straight} : B)$
  $(\textit{flat}^? : (a : A) \to (\{s : S\} \to g\, s \equiv \mathsf{nothing} \to g\,(a \triangleleft s) \equiv \mathsf{nothing})\uplus$
                          $(\{s : S\} \to g\,(a \triangleleft s) \not\equiv \mathsf{nothing}))$
  $(\textit{nothing-from-e} : g\, e \equiv \mathsf{nothing})$
  $(\textit{jigsaw-condition} :$
    $\{a : A\}\, \{b : B\}\, \{s\, s' : S\} \to$
    $(g\, s \equiv \mathsf{just}\,(b,\, s'))\uplus ((g\, s \equiv \mathsf{nothing}) \times (g\,(a \triangleleft s) \not\equiv \mathsf{nothing}) \times (b \equiv \textit{straight}) \times (s' \equiv s)) \to$
    **let** $(b',\, a') = \textit{piece}\,(a,\, b)$ **in** $g\,(a \triangleleft s) \equiv \mathsf{just}\,(b',\, a' \triangleleft s'))$
  **where**

  $\textit{fill} : \{s : S\}\, (a : A) \to \mathsf{CoalgList}\, B\, g\, s \to \mathsf{CoalgList}\, B\, g\,(a \triangleleft s)$
  $\textit{decon}\,(\textit{fill}\, a\, bs)$ **with** $\textit{decon}\, bs$
  $\textit{decon}\,(\textit{fill}\, a\, bs)\,\mid\, \langle\, eq\,\rangle$ **with** $\textit{flat}^?\, a$
  $\textit{decon}\,(\textit{fill}\, a\, bs)\,\mid\, \langle\, eq\,\rangle\,\mid\, \mathsf{inj}_1\,\textit{flat}\quad\ \ = \langle\,\textit{flat}\, eq\,\rangle$
  $\textit{decon}\,(\textit{fill}\, a\, bs)\,\mid\, \langle\, eq\,\rangle\,\mid\, \mathsf{inj}_2\,\textit{not-flat} =$
    **let** $(b',\, a') = \textit{piece}\,(a,\, \textit{straight})$
    **in** $b' :: \langle\,\textit{jigsaw-condition}\,(\mathsf{inj}_2\,(eq,\, \textit{not-flat},\, \mathsf{refl},\, \mathsf{refl}))\,\rangle\rangle\,\textit{fill}\, a'\, bs$
  $\textit{decon}\,(\textit{fill}\, a\, bs)\,\mid\, b :: \langle\, eq\,\rangle\, bs' =$
    **let** $(b',\, a') = \textit{piece}\,(a,\, b)$
    **in** $b' :: \langle\,\textit{jigsaw-condition}\,(\mathsf{inj}_1\, eq)\,\rangle\,\textit{fill}\, a'\, bs'$

  $\textit{jigsaw} : \{s : S\} \to \mathsf{AlgList}\, A\,(\triangleleft)\, e\, s \to \mathsf{CoalgList}\, B\, g\, s$
  $\textit{decon}\,(\textit{jigsaw}\,[\,]) = \langle\,\textit{nothing-from-e}\,\rangle$
  $\textit{jigsaw}\,(a :: as)\quad = \textit{fill}\, a\,(\textit{jigsaw}\, as)$

■ **Figure 10**  Metamorphisms in the general (possibly finite) jigsaw model (with the vertical placement strategy)

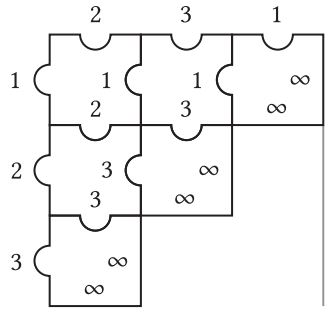

If we performed some case analysis at goal 5 like for goal 4, we might have been able to come up with an algorithm that could decrease column lengths when going left, saving more jigsaw pieces (although probably not for heapsort).

## About the author

**Hsiang-Shang Ko** is an Assistant Research Fellow at the Institute of Information Science, Academia Sinica, Taiwan. Contact him at joshko@iis.sinica.edu.tw.

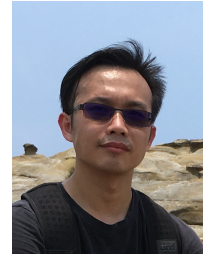